\PassOptionsToPackage{numbers,sort&compress}{natbib}

\documentclass[unnumsec,webpdf,contemporary,large,nocrop,noinfo]{oup-authoring-template}

\usepackage{publication-style}

\begin{document}

\journaltitle{Briefings in Bioinformatics}
\DOI{https://doi.org/10.1093/bib/bbag365}
\copyrightyear{2026}
\pubyear{2026}
\volume{27}
\vol{27}
\issue{4}
\appnotes{Review}
\firstpage{1}
\title{Imaging-anchored multiomics in cardiovascular disease: integrating cardiac imaging, bulk, single-cell, and spatial transcriptomics}

\author[1]{Minh H. N. Le}
\author[2]{Thanh-Huy Nguyen}
\author[3]{Tao Li}
\author[4]{Bao Quang Gia Le}
\author[5]{Han H. Huynh}
\author[4]{Monika Raj}
\author[3]{Carl Yang}
\author[2]{Min Xu}
\author[6,*]{Tuan Vinh}
\author[7,8,9,*]{Nguyen Quoc Khanh Le}

\address[1]{Section of Cardiovascular Medicine, Department of Internal Medicine, Yale School of Medicine, 333 Cedar Street, New Haven, CT 06510, United States}
\address[2]{Computational Biology Department, School of Computer Science, Carnegie Mellon University, 5000 Forbes Avenue, Pittsburgh, PA 15213, United States.}
\address[3]{Department of Computer Science, Emory University, 201 Dowman Drive, Atlanta, GA 30322, United States}
\address[4]{Department of Chemistry, Emory University, 1515 Dickey Drive, Atlanta, GA 30322, United States}
\address[5]{Irell and Manella Graduate School of Biological Science, City of Hope, CA, United States}
\address[6]{Medical Sciences Division, University of Oxford, John Radcliffe Hospital, Headington, Oxford OX3 9DU, Oxfordshire, United Kingdom}
\address[7]{AIBioMed Research Group, Taipei Medical University, 250 Wu-Hsing Street, Xinyi District, Taipei 11031, Taiwan}
\address[8]{In-Service Master Program in Artificial Intelligence in Medicine, College of Medicine, Taipei Medical University, 250 Wu-Hsing Street, Xinyi District, Taipei 11031, Taiwan}
\address[9]{Translational Imaging Research Center, Taipei Medical University Hospital, 252 Wu-Hsing Street, Xinyi District, Taipei 11031, Taiwan}
\corresp[*]{Corresponding authors. Tuan Vinh, Medical Sciences Division, University of Oxford, John Radcliffe Hospital, Headington, Oxford OX3 9DU, Oxfordshire, United Kingdom. E-mail: \href{mailto:tuan.vinh@hertford.ox.ac.uk}{tuan.vinh@hertford.ox.ac.uk}; Nguyen Quoc Khanh Le, In-Service Master Program in Artificial Intelligence in Medicine, College of Medicine, Taipei Medical University, Taipei, Taiwan. E-mail: \href{mailto:khanhlee@tmu.edu.tw}{khanhlee@tmu.edu.tw}.}

\abstract{Cardiovascular disease arises from interactions between inherited risk, molecular programmes, and tissue-scale remodelling that are observed clinically through imaging. Cardiac MRI (CMR), computed tomography (CT), and echocardiography are integral to routine cardiovascular care, while bulk RNA sequencing, single-cell RNA sequencing, and spatial transcriptomics are providing increasingly detailed molecular characterization of cardiac tissue. Yet, these imaging and molecular data are still analysed in largely separate pipelines. This review examines joint representations that link cardiac imaging phenotypes to transcriptomic and spatially resolved molecular states. An imaging-anchored perspective is adopted in which echocardiography, CMR, and CT define a spatial phenotype of the heart, and bulk, single-cell and spatial transcriptomics provide cell-type- and location-specific molecular context. We define the representation requirements of each modality, compare multimodal fusion strategies, and synthesize integrative pipelines for radiogenomics, spatial alignment, and image-based gene-expression prediction, together with their validation requirements, limitations, and failure modes. Spatial multiomic maps of human myocardium and atherosclerotic plaque, together with single-cell, spatial, and multimodal medical foundation models, are advancing imaging-anchored multiomics; however, cost, scalability, and tissue availability remain substantial barriers to large-scale cardiovascular translation.}
\maketitle

\section{Introduction}

Cardiovascular disease (CVD) remains the leading global cause of mortality and disability despite major advances in prevention and treatment \cite{world_heart_statistics}. Guideline-directed workflows and classical risk models summarize a patient's state using a small number of imaging and clinical measurements that are robust but coarse.

At the same time, modern cardiology increasingly operates in a multiomics environment: bulk RNA-seq and proteomics are scalable to population-level cohorts, whereas sc/snRNA-seq \cite{litvivnukova2020cells} and spatial transcriptomics remain confined to smaller studies in which tissue availability \cite{tucker2020transcriptional}, sampling region, and processing infrastructure impose additional constraints.

This review focuses on \emph{imaging-anchored molecular phenotyping}.

Cardiac imaging describes anatomy and function in space and time, whereas transcriptomic readouts, from bulk RNA-seq through single-cell profiling to spatially resolved assays, capture the gene-expression programmes that drive or accompany structural change. These two data layers are still largely analysed in separate pipelines. Linking them could allow imaging features such as late gadolinium enhancement (LGE), regional strain, or plaque morphology to be interpreted in terms of cell states, pathways, and tissue niches, and may ultimately enable images to serve as non-invasive surrogates for molecular assays when biopsies or spatial omics are infeasible. We survey current computational approaches toward this goal, including radiogenomic association, spatial transcriptomics alignment, and image-based prediction of gene expression. Imaging is treated as the anchor modality for three practical reasons.

First, cardiac imaging is ubiquitous in clinical care, providing a shared coordinate system and a common language across sites.

Second, imaging captures the integrated result of genetics, haemodynamics, inflammation, and tissue remodelling, offering holistic but indirect readouts of biology.

Third, imaging can be registered, approximately in vivo and more precisely ex vivo, to histology and spatial transcriptomics, enabling cross-scale integration from voxel or segment to cell.

Several recent reviews survey machine learning for cardiac imaging \cite{leiner2019machine,ouyang2020_echonet_dynamic}, multimodal modelling for coronary artery disease, cardiovascular risk prediction \cite{zhou2025_multimodal_cad_review}, spatial technologies in cardiac biology \cite{spatial_heart_review_2025}, and medical foundation models that integrate imaging, text and clinical data \cite{moor2023_foundation_models,tu2024_generalist_biomed_ai}. Here, we position cardiac imaging as the primary spatial reference frame, spatial transcriptomics as the bridge from voxel to cell, and representation learning and multimodal fusion as the computational machinery for building joint latent spaces for mechanistic interpretation, disease subtyping, and cardiovascular translation.

\subsection{Scope and literature identification strategy}

This article is a narrative, methods-focused review rather than a systematic meta-analysis. To increase transparency and reproducibility, we used a structured identification strategy. Searches were performed across PubMed, Web of Science, Scopus, and IEEE Xplore, complemented by screening relevant preprint servers for recent methodological developments. Search terms were organized along four axes: (i) imaging modalities and phenotypes (echocardiography, CMR, CT/CTA, radiomics, fibrosis, strain, and plaque), (ii) molecular modalities (bulk RNA-seq, sc/snRNA-seq, spatial transcriptomics, and proteomics), (iii) integration tasks (radiogenomics, spatial registration/alignment, cross-modal prediction, or ``virtual'' transcriptomics) and (iv) computational paradigms (fusion strategies, representation learning, contrastive learning, graph neural networks (GNNs), transformers, and foundation models). Backward and forward citation chasing from exemplar studies and existing surveys identified additional relevant work, and seminal machine-learning papers were included where needed for technical completeness. We prioritized studies that (i) link imaging phenotypes to molecular readouts in cardiovascular settings, (ii) introduce multimodal or spatial integration methods directly applicable to imaging--omics problems, and/or (iii) provide clinically oriented validation relevant to biomarker discovery or risk prediction.

\subsection{Interdependent conditions and multimorbidity}

Cardiovascular phenotypes rarely occur in isolation. Heart failure frequently co-exists with atrial fibrillation, chronic kidney disease, diabetes, obesity, and inflammatory conditions, which can share upstream molecular drivers while altering imaging phenotypes through haemodynamic loading, neurohormonal signalling, and systemic inflammation. Imaging-anchored multiomics, therefore, benefits from modelling interdependent conditions rather than single labels: multi-task prediction, joint latent spaces shared across related endpoints, and causal frameworks that separate shared biology from condition-specific effects can reduce confounding and improve generalizability across heterogeneous clinical cohorts.

The remainder of the review is organized into four methodological layers, beginning with the modalities and biological context. Spatial transcriptomics bridges imaging-scale phenotypes and cellular molecular states, but it must span a fundamental resolution gap.

In vivo cardiac MRI and CT operate at millimetre-to-centimetre voxel scales, whereas current spatial transcriptomics platforms interrogate tissue at the micron-to-spot level: the widely used 10$\times$ Visium captures transcripts at 55 \textmu{}m spots with 100 \textmu{}m centre-to-centre spacing, next-generation platforms such as Visium HD and Stereo-seq achieve 2--8 \textmu{}m bin resolution approaching single-cell scale, and imaging-based systems such as Xenium and MERSCOPE resolve individual cells or subcellular compartments \cite{lim2025practical,pentimalli2025challenges}.

Bridging this gap requires deliberate experimental design at two levels. At the macroscopic level, the principal strategy is \emph{imaging-guided tissue sampling}: tissue blocks are dissected from regions predefined by in vivo imaging findings---such as LGE-positive versus LGE-negative myocardial segments, or histologically characterized infarct core, border zone, and remote myocardium---establishing a coarse but intentional correspondence between the imaging phenotype and the spatial molecular map \cite{kuppe2022_spatial_mi,calcagno2022_infarct_spatial}. Real-time CMR-guided endomyocardial biopsy can further improve targeting accuracy by directing sampling to LGE-defined lesions rather than relying on anatomical landmarks alone \cite{unterberg2017targeted,lurz2012diagnostic}. At the microscopic level, where ex vivo tissue is available, high-resolution ex vivo MRI or micro-CT serves as an intermediate registration bridge between in vivo imaging and the histology section on which spatial transcriptomics is performed. Despite these strategies, fine-grained, voxel-level alignment between in vivo imaging and spatial transcriptomic spots remains an open methodological challenge, limited by tissue deformation during sectioning, scale mismatch across modalities, and the absence of comprehensive spatial reference atlases of the human heart \cite{pentimalli2025challenges,kuppe2022_spatial_mi}.

\begin{figure*}[!t]
\centering
\includegraphics[width=\textwidth]{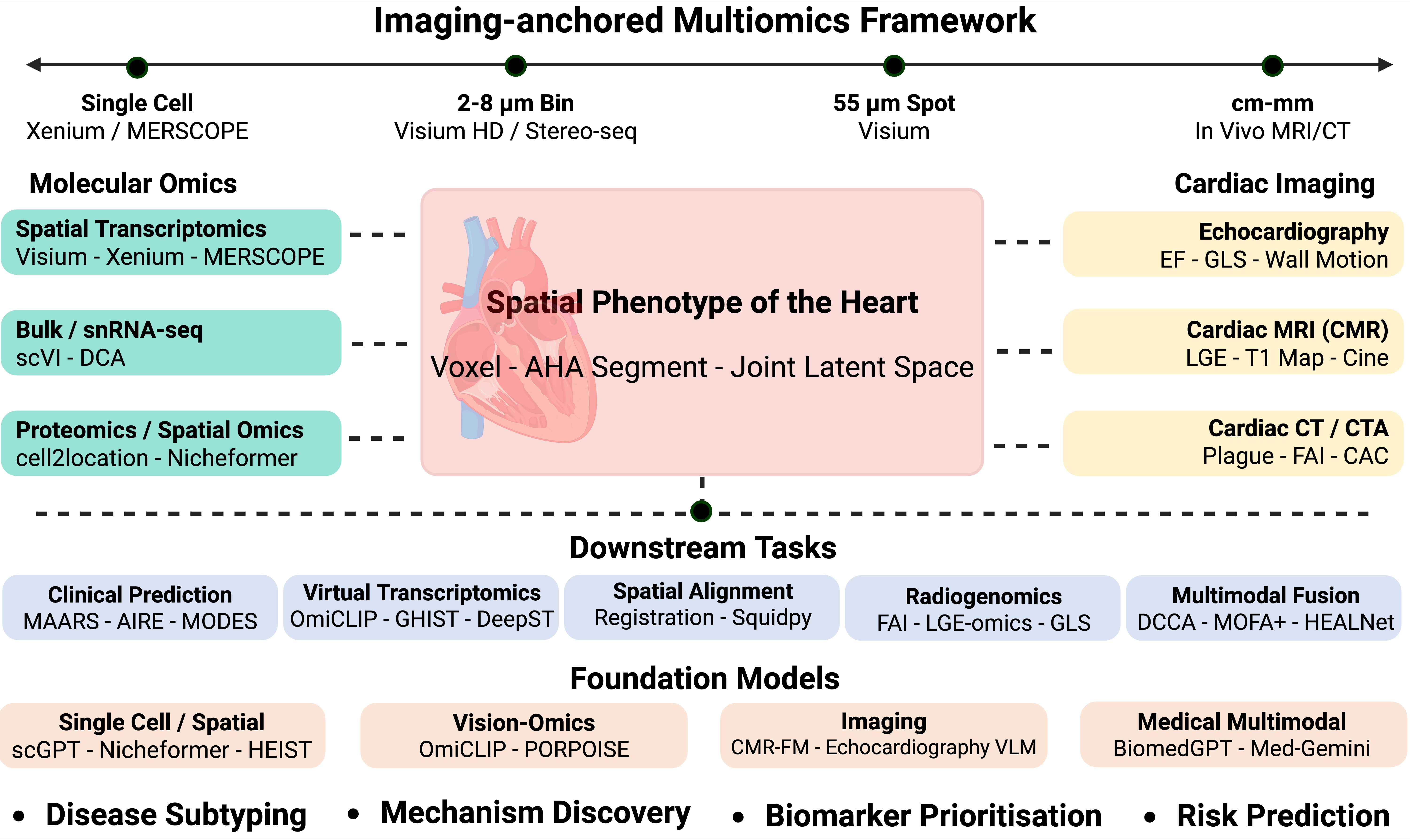}
\caption{The resolution scale illustrates the gap between in vivo cardiac imaging at centimetre-to-millimetre resolution and molecular spatial profiling, from Visium 55 \textmu{}m spots and Visium HD/Stereo-seq 2-8 \textmu{}m bins to single-cell platforms such as Xenium and MERSCOPE. Cardiac imaging modalities, including echocardiography, cardiac magnetic resonance imaging (CMR), and cardiac CT/CTA, define spatial phenotypes of the heart using measures such as ejection fraction, global longitudinal strain, wall motion, late gadolinium enhancement, T1 mapping, cine imaging, plaque features, fat attenuation index, and coronary artery calcium. These phenotypes can be indexed at the voxel, American Heart Association segment, or joint latent-space level and linked to molecular omics layers, including spatial transcriptomics, bulk or single-nucleus RNA sequencing, and proteomics/spatial omics. The resulting joint representations support downstream tasks including clinical prediction, virtual transcriptomics, spatial alignment, radiogenomics, and multimodal fusion. Foundation models spanning single-cell/spatial omics, vision--omics, imaging, and medical multimodal data provide computational support across the framework. Applications include disease subtyping, mechanism discovery, biomarker prioritisation, and risk prediction. Abbreviations: AHA, American Heart Association; CMR, cardiac magnetic resonance imaging; CT, computed tomography; CTA, computed tomography angiography; EF, ejection fraction; GLS, global longitudinal strain; LGE, late gadolinium enhancement; FAI, fat attenuation index; CAC, coronary artery calcium; scVI, single-cell variational inference; DCA, deep count autoencoder; DCCA, deep canonical correlation analysis; MOFA+, multi-omics factor analysis; VLM, vision--language model.}
\label{fig:imaging_anchored_framework}
\end{figure*}

Key public resources that combine cardiovascular imaging with molecular, waveform, or rich clinical data are summarized in Table~\ref{tab:multimodal_cardiology_datasets}, and representative imaging biomarkers that commonly serve as anchors (and their likely molecular correlates) are summarized in Table~\ref{tab:imaging_biomarkers}. The overall imaging-anchored multiomics framework is illustrated in Fig.~\ref{fig:imaging_anchored_framework}.

\begin{table*}[!tp]
\centering
\sffamily\fontsize{7.5bp}{9.45bp}\selectfont
\setlength{\tabcolsep}{1.5pt}
\renewcommand{\arraystretch}{1}
\caption{Public multimodal cardiology datasets relevant to imaging-anchored multiomics, including resources that directly pair cardiovascular imaging with molecular data and adjacent multimodal cardiovascular resources useful for encoder pretraining and transfer learning.}
\label{tab:multimodal_cardiology_datasets}
\begin{tabular}{@{}>{\raggedright\arraybackslash}p{\dimexpr 0.15645\textwidth - 2.500pt\relax}>{\raggedright\arraybackslash}p{\dimexpr 0.15090\textwidth - 2.500pt\relax}>{\raggedright\arraybackslash}p{\dimexpr 0.15090\textwidth - 2.500pt\relax}>{\raggedright\arraybackslash}p{\dimexpr 0.27576\textwidth - 2.500pt\relax}>{\raggedright\arraybackslash}p{\dimexpr 0.17809\textwidth - 2.500pt\relax}>{\raggedright\arraybackslash}p{\dimexpr 0.08790\textwidth - 2.500pt\relax}@{}}
\toprule
\textbf{Dataset/study} & \textbf{Type} & \textbf{Cardiovascular focus} & \textbf{Modalities included} & \textbf{Approximate scale and access} & \textbf{Link} \\
\midrule
UK Biobank imaging extension \cite{petersen2013imaging} &
Population imaging cohort &
Cardiac structure, function, and risk in mid-life adults &
CMR, vascular ultrasound and other MRI sequences linked to genome-wide genotyping, exome sequencing, plasma proteomics, metabolomics, EHR, and questionnaires &
$\sim$100,000 participants with CMR and >500,000 with genomics; application-based access &
\href{https://www.ukbiobank.ac.uk/}{UK Biobank} \\[1.5pt]
Multi-Ethnic Study of Atherosclerosis (MESA) \cite{bild2002_mesa} &
Population cohort with multimodal imaging &
Subclinical atherosclerosis and incident CVD in a diverse community sample &
CMR, coronary artery calcium CT, carotid ultrasound, and other imaging linked to genotype, labs, and risk-factor profiling &
$\sim$6,800 participants; controlled access via NHLBI repositories &
\href{https://biolincc.nhlbi.nih.gov/studies/mesa/}{BioLINCC} \\[1.5pt]
Hypertrophic Cardiomyopathy Registry (HCMR) \cite{kramer2015_hcmr} &
Disease registry and biobank &
Risk stratification in hypertrophic cardiomyopathy &
CMR, echocardiography and electrocardiography (ECG) combined with targeted gene sequencing and circulating biomarkers &
$\sim$2,700 patients; collaborative access via data-use agreements &
\href{https://www.hcmregistry.org/}{HCMR} \\[1.5pt]
MIMIC-IV and related PhysioNet modules \cite{gow2023_mimic_iv_ecg} &
Critical-care waveform and EHR resource &
Acute CVD and haemodynamic instability in ICU/ED settings &
Twelve-lead ECG waveforms linked to high-resolution EHR; additional echo and note modules for subsets &
$\sim$800,000 ECGs in $\sim$160,000 subjects; credentialed access via PhysioNet &
\href{https://physionet.org/content/mimiciv/3.1/}{PhysioNet} \\[1.5pt]
EchoNet family (EchoNet-Dynamic, EchoNet-LVH) \cite{ouyang2020_echonet_dynamic,duffy2022high} &
Curated echocardiography video datasets &
LV systolic function, wall motion, and hypertrophy phenotypes &
Apical four-chamber echo videos with contour tracings, volumetric measurements, and labels (EF, LVH) &
On the order of $10^{4}$ videos; de-identified datasets downloadable for research &
\href{https://echonet.github.io/dynamic/}{EchoNet} \\[1.5pt]
Cardiac Atlas Project CMR cohorts \cite{fonseca2011_cardiac_atlas} &
Aggregated CMR repositories &
Cardiac structure and function in health and early disease &
3D cine CMR studies with segmentations, finite-element meshes and clinical covariates in selected cohorts &
Thousands of CMR exams; access under data-use agreements &
\href{https://www.cardiacatlas.org/}{Cardiac Atlas} \\[1.5pt]
Spatial multi-omic map of human myocardial infarction \cite{kuppe2022_spatial_mi} &
Human tissue atlas with single-cell and spatial multiomics &
Early and late remodelling after acute myocardial infarction &
snRNA-seq, snATAC-seq, Visium spatial transcriptomics, and matched histology across infarct core, border zone, and remote myocardium &
31 samples from 23 patients; processed data via public portals and archives &
\href{https://data.humancellatlas.org/}{HCA portal} \\[1.5pt]
Coronary plaque spatial transcriptomics studies \cite{campos2025_spatial_plaque,sun2023spatial,gastanadui2024_coronary_spatial_plaque} &
Plaque-level spatial-omics studies &
Plaque stability, inflammatory niches, and high-risk morphologies &
Spatial transcriptomics of human coronary arteries with histology and plaque morphology; some cohorts include CT or invasive imaging &
Dozens of plaques; expression matrices; and metadata commonly deposited to GEO/ENA (see articles) &
\href{https://www.ncbi.nlm.nih.gov/geo/}{GEO} \\[1.5pt]
\bottomrule
\end{tabular}
\end{table*}

\begin{table*}[!tp]
\centering
\sffamily\fontsize{7.5bp}{9.45bp}\selectfont
\setlength{\tabcolsep}{1.5pt}
\renewcommand{\arraystretch}{1}
\caption{Representative cardiovascular imaging biomarkers that often serve as anchors in imaging-anchored multiomics, with typical computational extraction strategies and plausible molecular correlates.}
\label{tab:imaging_biomarkers}
\begin{tabular}{@{}>{\raggedright\arraybackslash}p{\dimexpr 0.22271\textwidth - 2.250pt\relax}>{\raggedright\arraybackslash}p{\dimexpr 0.23880\textwidth - 2.250pt\relax}>{\raggedright\arraybackslash}p{\dimexpr 0.23880\textwidth - 2.250pt\relax}>{\raggedright\arraybackslash}p{\dimexpr 0.29969\textwidth - 2.250pt\relax}@{}}
\toprule
\textbf{Biomarker (modality)} & \textbf{Clinical interpretation} & \textbf{Typical computational extraction} & \textbf{Plausible or reported molecular correlates} \\
\midrule
LGE extent and pattern (CMR) \cite{kuruvilla2014_lge_meta} &
Focal myocardial scar/fibrosis; arrhythmic and HF risk &
Segmentation of LGE-positive regions; regional burden by AHA segments; texture/radiomics &
Fibroblast activation, collagen deposition (ECM programmes), immune infiltration in border zones; stress and remodelling programmes in cardiomyocytes \\[1.5pt]
Global longitudinal strain (echo) \cite{kalam2014_gls_meta} &
Subclinical systolic dysfunction; prognostic across many CVD states &
Speckle-tracking strain curves; regional strain maps; video encoders for deformation fields &
Energetic stress, sarcomeric dysfunction, inflammatory signalling; chamber-specific cardiomyocyte stress responses \\[1.5pt]
Low-attenuation noncalcified plaque (coronary CTA) \cite{williams2020_scotheart_lap} &
Lipid-rich plaque; higher MI risk &
Plaque segmentation and attenuation profiling; radiomics; phenotype classifiers &
Macrophage-driven inflammation, necrotic core formation, matrix remodelling; immune niches at plaque shoulders \\[1.5pt]
Napkin-ring sign/high-risk plaque morphology (CTA) \cite{otsuka2013_napkin_ring} &
High-risk plaque architecture; potential plaque vulnerability &
CNN/ViT-based plaque morphology classification; handcrafted geometry features &
Protease and cytokine signalling, macrophage activation, smooth muscle cell phenotypic switching; local hypoxia \\[1.5pt]
Perivascular fat attenuation index (FAI) (CTA) \cite{oikonomou2018_pv_fai_risk} &
Non-invasive readout of coronary inflammation &
Perivascular fat segmentation; attenuation gradients; radiotranscriptomic mapping &
Inflammatory adipose and vascular wall transcripts; cytokine signalling and fibrosis-related programmes \\[1.5pt]
\bottomrule
\end{tabular}
\end{table*}

\section{Foundations: modalities and biological context}

\subsection{Cardiac imaging as a spatial phenotype}

Cardiac imaging provides a high-level description of heart structure and function that is naturally indexed in space (and often time). Echocardiography offers real-time 2D and 3D views at the bedside, and resources such as EchoNet-Dynamic and EchoNet-LVH have accelerated deep learning for view classification, segmentation, and disease detection in echocardiographic videos \cite{ouyang2020_echonet_dynamic,duffy2022high}. CMR provides cine imaging for ventricular function, parametric mapping for tissue characterization, and LGE for focal fibrosis; population-scale examples are available in the UK Biobank programme \cite{petersen2013imaging}. Cardiac CT, particularly coronary CT angiography, visualizes coronary anatomy, and plaque morphology, including low-attenuation plaque and calcification patterns associated with risk \cite{ferencik2018use}.

These modalities define a spatial phenotype: voxels or segments correspond to physical locations and encode tissue composition, perfusion, and mechanical function. Traditional analyses compress this into global or regional summary metrics, but imaging--omics integration often benefits from intermediate representations that retain spatial detail \cite{radiomics_cmr_2020}. These imaging-derived features provide the scaffold onto which molecular context can be mapped.

\subsection{Imaging biomarkers as anchors and their molecular correlates}

Many imaging--omics studies begin with clinically established biomarkers and then ask which molecular programmes best explain their variation. Table~\ref{tab:imaging_biomarkers} summarizes representative biomarkers used as anchors and plausible cellular and pathway-level correlates.

\subsection{Omic layers: bulk, single-cell, and proteomic context}

Transcriptomic and related omic measurements capture dynamic molecular states shaped by genetic background and environment. Bulk RNA-seq from biopsies or explanted myocardium profiles pathways active in cardiomyopathies, myocarditis, and heart failure \cite{koenig2022_cardiomyopathy_rnaseq}, while plasma proteomics quantifies circulating inflammatory, fibrotic, and metabolic proteins that can predict events beyond classical risk factors \cite{ganz2016_protein_risk_jama}. These bulk measurements average over cell types and niches, offering global molecular context but limited resolution.

Single-cell and single-nucleus RNA sequencing provide cell-level resolution and reference atlases of cardiomyocytes, fibroblasts, endothelial and smooth muscle cells, pericytes, and immune populations \cite{litvivnukova2020cells,wang2020_hf_scseq}. In disease, cardiomyocytes often converge onto stress programmes, whereas fibroblasts, endothelial cells, and immune cells diversify into disease-associated states enriched for inflammatory and fibrotic genes \cite{tucker2020transcriptional,koenig2022_cardiomyopathy_rnaseq}. Although this review emphasizes transcriptomic layers, additional omics (epigenomics, metabolomics, and proteomics) are frequently available and can be incorporated via multi-view factor models or graph-based encoders.

\subsection{Spatial transcriptomics as a bridge}

Spatial transcriptomics preserves where transcripts are located within tissue. Cardiovascular studies have mapped infarct cores, border zones, and remote myocardium \cite{kuppe2022_spatial_mi}, delineated microdomains in cardiomyopathies \cite{lee2025_spatial_hf}, and characterized vulnerable plaque regions in atherosclerotic arteries \cite{campos2025_spatial_plaque}. Spatial clustering and domain discovery are central for revealing microenvironments and interaction niches, and are typically coupled to marker-gene analysis and deconvolution against single-cell references---using methods such as cell2location \cite{kleshchevnikov2022cell2location}---to infer cell-type proportions within each capture spot \cite{kuppe2022_spatial_mi}.

For imaging-anchored multiomics, spatial transcriptomics is particularly attractive because it resides close to histology and can be warped toward imaging coordinate systems via intermediate registration steps. Spot-level expression, deconvolved cell composition, and histology features together provide a molecular annotation of tissue architecture that can be linked to imaging-derived phenotypes.

\section{Representation learning across modalities}

\subsection{Imaging encoders}

Cardiac imaging data are high-dimensional, often 3D and sometimes time-resolved. Representation learning aims to compress these data into compact embeddings that preserve clinically and biologically relevant information. For imaging-anchored multiomics, the encoder must satisfy two requirements that go beyond standard segmentation or classification: first, it must produce embeddings at spatial granularities that can be aligned with molecular data---from whole-heart summaries matched to bulk transcriptomics, through regional American Heart Association segments paired with biopsy-level profiles, to voxel-level features co-registered with individual spatial transcriptomic spots. Second, it must learn generalizable representations from large unlabelled imaging archives, because the paired imaging--omics samples available for joint training typically number in the tens to low hundreds.

\subsubsection{Convolutional architectures}

Convolutional neural networks (CNNs) and their 3D or spatiotemporal variants remain the workhorses for echocardiography and CMR segmentation, view classification and disease prediction \cite{leiner2019machine,ouyang2020_echonet_dynamic}. Population-scale deployment on the UK Biobank, where fully convolutional networks have segmented cardiac chambers in tens of thousands of subjects \cite{bai2018automated}, illustrates that CNN-derived phenotypes---volumes, mass, wall thickness---can already serve as high-throughput imaging features for genetic association \cite{bai2020population} and, by extension, for radiogenomic studies that link imaging variation to transcriptomic programmes---as demonstrated by radiotranscriptomic mapping of perivascular fat \cite{oikonomou2019novel}.

For cine-MRI and echocardiographic videos, 3D CNNs or 2D CNNs combined with temporal convolution or recurrent layers can model wall motion and valve dynamics \cite{leiner2019machine,ouyang2020_echonet_dynamic}. Crucially for multiomics integration, intermediate convolutional feature maps retain spatial layout: region- or voxel-level embeddings can be extracted from these maps and paired with molecular measurements at matching anatomical locations \cite{radiomics_cmr_2020}, providing a direct bridge between the imaging encoder, and the spatial or regional molecular annotations discussed in the preceding section.

\subsubsection{Transformer and self-supervised architectures}

Vision Transformers (ViTs) \cite{dosovitskiy2021_vit} and masked autoencoders (MAEs) \cite{he2022_mae} offer a complementary design. Images or volumes are decomposed into patches, projected into tokens, and processed via self-attention, which captures long-range dependencies across remote myocardial or coronary segments. This property matters when the imaging phenotype of interest---such as diffuse fibrosis detectable on T1 mapping, global longitudinal strain, or multi-territory plaque burden---reflects spatially distributed biology that local convolutions may underrepresent. Self-supervised objectives such as masked patch prediction or contrastive learning on large unlabeled archives enable these encoders to learn generic cardiac representations that transfer to downstream tasks with comparatively few labels: a CMR foundation model pretrained on 36 million images \cite{jacob2025towards} and a vision--language echocardiography model trained on over one million video--report pairs \cite{christensen2024vision} both demonstrate strong few-shot and zero-shot transfer to clinical cardiac tasks, respectively. For imaging--omics integration, such pretrained encoders can provide high-quality, spatially structured embeddings without consuming the scarce paired imaging--molecular samples for encoder training; those paired data can instead be reserved for the fusion and alignment layers discussed in the ``Multimodal fusion: methods, pipelines, and evaluation'' section. Bridging from imaging to molecular states at spot or cell resolution, however, requires a different class of encoder that jointly learns from histology and transcriptomics. Vision--omics foundation models such as OmiCLIP, trained on 2.2 million paired haematoxylin and eosin (H\&E)--transcriptome patches across 32 organs including heart failure tissue \cite{chen2025visual}, can predict spatial gene expression and infer cell-type composition directly at Visium spot level. Similarly, self-supervised ViTs trained on large histology archives have been shown to predict and spatially localize individual transcript abundances across 23 human tissue types \cite{cisternino2024self}. Although these histology--omics models have been demonstrated primarily in oncology, the architectures are directly adaptable to cardiac tissue sections once sufficiently large cardiac spatial datasets become available---a prospect discussed in the ``Future directions'' section. Training patient-level imaging--omics models end-to-end from scratch on small paired cohorts meanwhile risks severe overfitting to confounders such as site or scanner batch, a failure mode discussed in the ``Bottlenecks, root causes, and failure modes'' section.

\subsubsection{From handcrafted radiomics to learned features}

Before deep encoders became dominant, handcrafted radiomics---shape, intensity histogram, and texture descriptors extracted from segmented regions of interest---provided the primary featurization of cardiac images for association studies \cite{radiomics_cmr_2020}. Radiomics remains valuable for interpretability: first-order and texture features have transparent physical meanings and can be mapped onto tissue-level processes such as fibrosis heterogeneity or myocardial oedema \cite{radiomics_cmr_2020}. In practice, hybrid strategies that concatenate radiomics with deep embeddings often outperform either alone, because handcrafted features capture domain priors that a data-hungry deep model may miss in small paired cohorts. For imaging--omics integration specifically, radiomic features offer the further advantage that they can be extracted per AHA segment or per plaque, yielding a natural correspondence with region-level molecular measurements from bulk RNA-seq or deconvolved spatial transcriptomics.

\subsubsection{Multi-scale embeddings for molecular alignment}

The choice of embedding scale should match the granularity of available molecular data and the biological question. Global (whole-heart) embeddings suffice for patient-level radiogenomic association, where the aim is to correlate an imaging-derived trait such as left ventricular mass with bulk gene-expression or plasma proteomic modules. Regional embeddings at the level of AHA segments or individually segmented plaques are appropriate for linking to biopsy-level profiles or cell-type compositions estimated by deconvolution methods such as cell2location \cite{kleshchevnikov2022cell2location}. Voxel-level features are needed to map gene-expression gradients across infarct border zones or plaque shoulders where spatial transcriptomics provides spot-level resolution \cite{kuppe2022_spatial_mi}. Designing encoders that can output embeddings at multiple scales simultaneously, or that can be queried at arbitrary coordinates via spatial interpolation, remains an open challenge that directly determines the resolution at which imaging and molecular data can be fused.

\subsection{Omics and single-cell encoders}

Bulk transcriptomic data are typically high-dimensional, with thousands of genes, substantial measurement noise, and pervasive gene--gene co-expression organized into modules of functionally related genes \cite{langfelder2008wgcna}, which makes the data amenable to dimensionality reduction.

Standard analysis toolkits---Seurat \cite{hao2024dictionary} in R and Scanpy \cite{wolf2018scanpy} in Python---implement principal component analysis (PCA)-based dimensionality reduction, graph-based clustering (Leiden/Louvain), differential expression testing and visualization via UMAP or t-SNE, providing the default entry point for both bulk and single-cell workflows. For deeper denoising, DCA \cite{eraslan2019single} trains a zero-inflated negative binomial autoencoder to impute dropout events in count matrices. scVI \cite{lopez2018_scvi}, implemented within the scvi-tools ecosystem \cite{gayoso2022_scvi_tools}, uses a variational autoencoder to jointly perform dimensionality reduction, batch correction, and differential expression in a single probabilistic framework. MOFA+ \cite{argelaguet2018_mofa} decomposes variation across multiple omic layers into shared factors capturing cross-modality biology and private factors specific to one layer. For imaging--omics integration, the low-dimensional representations produced by these tools---whether PCA components, scVI latent variables, or MOFA+ shared factors---serve as molecular summaries that can be aligned with imaging embeddings.

Single-cell transcriptomics introduces additional complexity: sparse count data, batch effects, and large numbers of cells. The scVI framework described above was designed specifically for these challenges, and its latent spaces support cell clustering, trajectory inference, and integration across donors and conditions \cite{lopez2018_scvi}.

Beyond dataset-specific latent models, single-cell, and multiomics foundation models trained on tens of millions of cells are beginning to appear \cite{singlecell_foundation_2024,nicheformer_2025}. Large-scale transformers and state-space models that ingest gene-expression vectors, chromatin accessibility, spatial coordinates, or perturbation labels can provide transferable embeddings that generalize across tissues and species \cite{singlecell_foundation_2024,nicheformer_2025}. Early evaluations suggest that cardiac and vascular cell states are represented faithfully in such models, which raises the prospect of using pretrained single-cell foundation models as generic encoders in cardiovascular multi-omics.

\subsection{Spatial transcriptomics encoders}

Spatial transcriptomics data reside on grids of capture spots or irregular spatial point sets. Unlike scRNA-seq, where each observation represents a single cell, sequencing-based platforms such as Visium assign each spot a mixture of transcripts from multiple co-located cells \cite{lim2025practical}, and even higher-resolution platforms retain spatial dependencies absent from dissociated-cell data. Applying standard scRNA-seq dimensionality reduction---PCA followed by k-nearest-neighbour graphs constructed purely in expression space---to spots ignores both this within-spot cellular heterogeneity and the tissue-level spatial structure, motivating dedicated spatial encoders.

\subsubsection{Graph-based spatial domain identification}

GNN methods address this by constructing graphs in which nodes represent spots and edges connect spatial neighbours, so that message passing yields embeddings that respect both expression similarity and physical adjacency. SpaGCN \cite{hu2021_spagcn} integrates gene expression, spatial coordinates, and histology via graph convolution to identify spatial domains with coherent expression and morphology, and is used for spatially variable gene detection. BayesSpace \cite{zhao2021spatial} adopts a Bayesian framework that models spatial neighbourhood priors to achieve sub-spot resolution clustering, enabling finer domain delineation than the native Visium grid. STAGATE \cite{dong2022deciphering} introduces an adaptive graph attention auto-encoder that learns the similarity of neighbouring spots through attention weights, improving boundary detection between spatial domains and supporting denoising of expression data. GraphST \cite{long2023spatially} combines GNNs with self-supervised contrastive learning---minimizing embedding distance between spatially adjacent spots while maximizing it for non-adjacent ones---and is the first method to jointly perform spatial clustering, multi-sample batch integration and cell-type deconvolution in a unified framework. Spatially informed graph attention architectures such as STMSGAL \cite{zhou2025_stmsgal} further highlight how multi-scale attention and graph signal processing can be combined to improve spatial domain identification in noisy settings.

\emph{Multimodal histology--expression encoders.} When paired histology images are available, multimodal encoders that jointly process H\&E patches and spot-level gene-expression vectors can learn richer representations than expression-only methods. Contrastive frameworks such as mclSTExp \cite{min2024multimodal} align image features from a DenseNet encoder with spot expression and positional embeddings from a Transformer encoder in a shared latent space, and use the learned alignment to predict spatial gene expression from H\&E images alone. A recent benchmark of 11 histology-to-expression prediction methods confirmed that such H\&E-based virtual transcriptomics is feasible across tissue types, while identifying batch effects and limited cross-dataset generalizability as key remaining challenges \cite{wang2025benchmarking}. Foundation-scale architectures trained on hundreds of thousands of spots from diverse tissues---including graph-based models such as HEIST \cite{heist_2025} and cross-attention or diffusion approaches such as DeepST \cite{deepst_2022}---push this further by learning transferable spatial representations in a self-supervised regime.

\subsubsection{Relevance to imaging--omics integration}

For imaging-anchored multiomics, these spatial encoders serve two complementary roles. First, they compress high-dimensional spot-level expression into low-dimensional spatial feature maps that can be co-registered to in vivo imaging via the alignment pipelines described in the ``Spatial molecular alignment and tissue--image correspondence'' section. Second, imaging encoders can be designed to output features at the resolution of projected spatial spots---for instance by resampling CMR-derived LGE intensity, T1 values or strain at spot coordinates---enabling feature-level pairing between imaging and molecular embeddings within a shared graph or latent space. Although most current spatial encoder demonstrations are in oncology, the architectures should be evaluated in cardiac and vascular tissue as cardiovascular spatial atlases expand (Future directions).

\subsection{Self-supervision, batch effects, and limitations}

Across modalities, self-supervised learning is becoming the dominant pretraining paradigm, driven by the observation that labelled and paired imaging--omics data are scarce, whereas unimodal archives are large. For cardiac imaging, contrastive and masked-modelling objectives are now producing encoders that rival or exceed supervised baselines in label-limited regimes. In echocardiography, a self-supervised segmentation pipeline trained on only 450 studies without any manual chamber labels produced cardiac chamber measurements across >18,000 echocardiograms that matched both inter-clinician variability and fully supervised learning \cite{ferreira2025selfsupervised}. At the CMR scale, the foundation model of Jacob et al. \cite{jacob2025towards}, pretrained with masked image modelling on 36 million CMR images, transfers to segmentation and function quantification with minimal labelled data, and the vision--language echocardiography model of Christensen et al. \cite{christensen2024vision} achieves zero-shot view classification from over one million video--report pairs. Together, these models illustrate that self-supervised cardiac imaging encoders can now produce high-quality embeddings without consuming the scarce paired imaging--molecular samples needed for downstream fusion.

For transcriptomics, analogous self-supervised strategies are maturing rapidly. Masked gene modelling, in which a fraction of gene-expression values are withheld and the model is trained to reconstruct them, underpins single-cell foundation models such as scGPT, a generative pretrained transformer trained on over 33 million cells that supports batch correction, cell-type annotation, and perturbation prediction after fine-tuning \cite{cui2024scgpt}, and scFoundation \cite{singlecell_foundation_2024}. Denoising autoencoders such as DCA \cite{eraslan2019single} and variational frameworks such as scVI \cite{lopez2018_scvi} capture co-expression structure without labels and remain standard preprocessing steps in single-cell pipelines.

Cross-modal self-supervised objectives further enrich representations for integration. A cross-modal autoencoder trained on paired electrocardiography, CMR, and clinical variables at biobank scale demonstrated that a shared latent space can simultaneously support imaging-based prediction, genome-wide association, and counterfactual simulations \cite{crossmodal_cardiac_ae_2023}, and CLIP-style contrastive objectives that align imaging--text pairs in radiology \cite{radford2021_clip,zhang2022_convirt} provide a template for imaging--omics contrastive pretraining when paired cohorts become available.

In parallel, multimodal medical foundation models that couple imaging, free-text reports, and other clinical data have been proposed for radiology and general biomedical tasks \cite{moor2023_foundation_models,tu2024_generalist_biomed_ai,foundation_radiology_2025,sun2024_medical_mmfm}. These architectures typically combine masked modelling and contrastive objectives across modalities and offer a natural starting point for incorporating omic inputs once suitably annotated cohorts are available. Recent large vision--language models for radiology and generalist biomedical AI further underline the need for careful curation, auditing, and domain-specific fine-tuning when such backbones are repurposed for multimodal bioinformatics \cite{foundation_radiology_2025,strotzer2024_gpt4v_radiology}.

Batch effects and cross-centre variability pose persistent challenges that affect both modalities and their joint analysis. Imaging domain shifts arise from differences in scanners, acquisition protocols, and reconstruction algorithms; a recent survey of deep-learning-based MRI harmonization catalogued a rapidly growing toolkit of GANs, VAEs, normalizing flows, and disentangled-representation methods for reducing scanner-specific bias, while noting that most approaches still lack standardized evaluation on cardiac data \cite{abbasi2024deep}. For omics, batch-correction methods such as ComBat \cite{johnson2007_combat} and Harmony \cite{korsunsky2019_harmony}, and joint latent models such as scVI \cite{lopez2018_scvi}, are well established; however, harmonizing across fundamentally different platforms---for instance, Visium versus Visium HD or 10$\times$ Chromium versus Smart-seq---remains challenging even with these tools. When imaging and omics batches are confounded, as often happens when specific imaging centres also provide the only tissue samples, the risk of learning spurious associations increases sharply. A practical limitation is sample size: paired imaging--omics datasets remain small, often involving tens to hundreds of patients, which compounds the risk of overfitting and confounding by site or protocol. Representation-learning strategies must, therefore, be coupled to stringent validation and sensitivity analyses, including leave-one-site-out designs and negative controls with permuted pairings, as discussed in the ``Bottlenecks, root causes, and failure modes'' section.

\section{Multimodal fusion: methods, pipelines, and evaluation}

\begin{figure*}[!t]
\centering
\includegraphics[width=\textwidth]{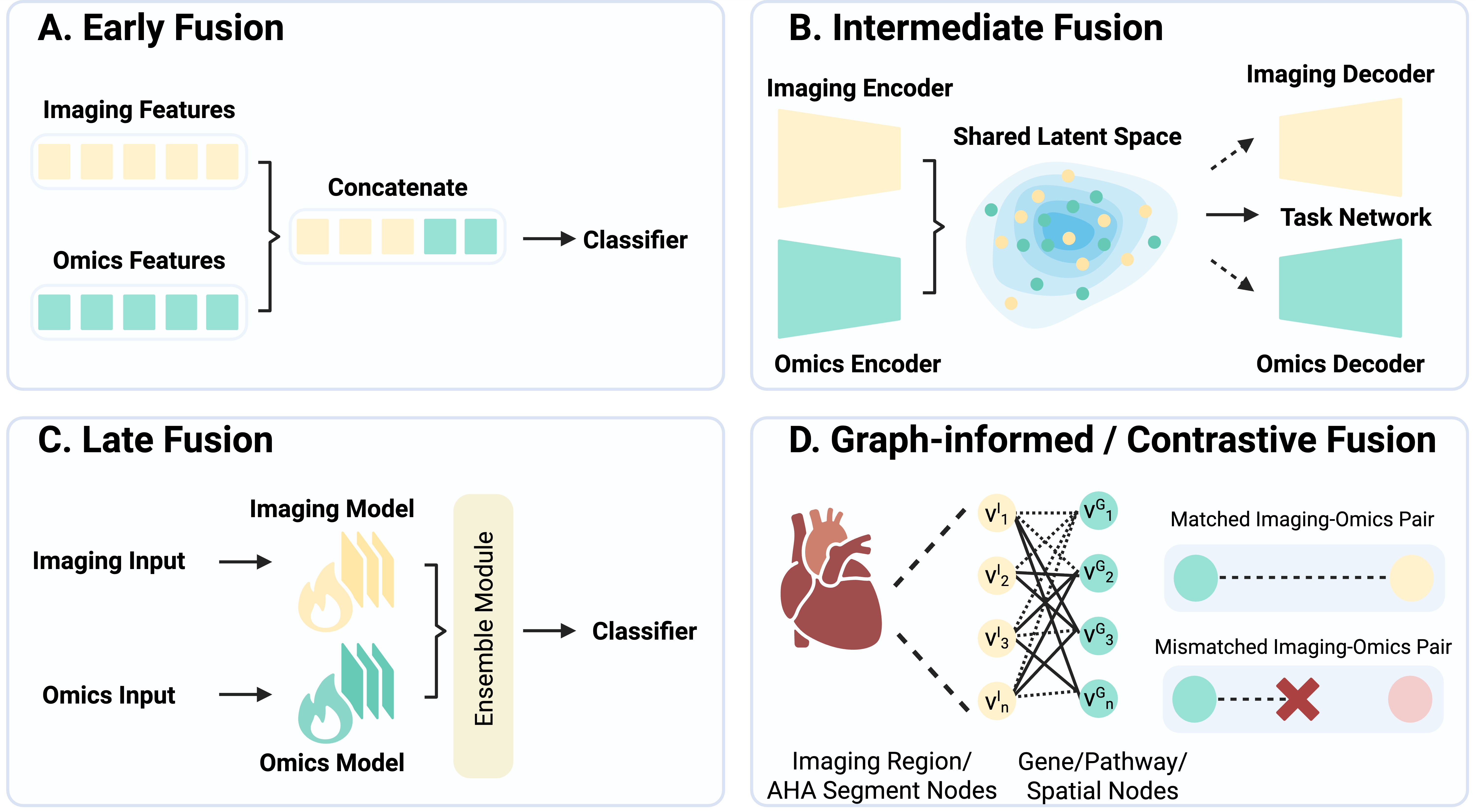}
\caption{Multimodal fusion architectures for radiomics--omics integration. Panel (A): early fusion concatenates radiomics and omics feature vectors followed by a single predictive model. Panel (B): intermediate fusion uses modality-specific encoders and a shared latent space, often with reconstruction heads; the shared latent captures biology concordant across modalities while private components retain modality-specific information. Panel (C): late fusion combines predictions from separate radiomics and omics models via an ensemble or calibration layer; it is robust to missing modalities but cannot capture cross-modal feature interactions. Panel (D): graph-based and contrastive approaches construct cross-modal latent spaces using biological or spatial graphs and contrastive objectives.}
\label{fig:multimodal_fusion}
\end{figure*}

This section examines how radiomic, transcriptomic, and spatial transcriptomic representations can be combined in joint models. We use ``fusion'' broadly to include architectures, learning objectives, and evaluation frameworks; Fig.~\ref{fig:multimodal_fusion} summarizes early, intermediate, late, and graph-based strategies.

A prerequisite to any fusion strategy is understanding the data generation pipeline. In clinical cardiovascular practice, echocardiography, CMR, and CT are acquired routinely and cover the entire heart or vascular territory at millimetre-to-centimetre resolution. Radiomics features---shape, intensity histogram, texture, and deep-learned descriptors extracted from segmented regions of interest \cite{radiomics_cmr_2020,hassani2020myocardial}---constitute the imaging-derived layer of the multiomics stack. Molecular profiling, by contrast, requires physical tissue: endomyocardial biopsies, surgical specimens, explanted hearts or, for circulating analytes, blood draws. Because tissue is scarce and costly, radiomics features play a critical upstream role in defining which regions will be profiled molecularly. LGE-positive versus LGE-negative myocardial segments, infarct border zones delineated on parametric maps, or coronary plaque regions classified by CT attenuation can all serve as radiomics-defined targets that guide biopsy or tissue-block dissection \cite{kuppe2022_spatial_mi,calcagno2022_infarct_spatial}. Real-time CMR-guided endomyocardial biopsy can further improve targeting accuracy by directing sampling to LGE-defined lesions rather than relying on anatomical landmarks alone \cite{unterberg2017targeted,lurz2012diagnostic}. The resulting molecular data---bulk RNA-seq, sc/snRNA-seq, spatial transcriptomics, plasma proteomics---are then linked back to the radiomics features from which the sampling decision was derived, forming the radiogenomic pairs that fusion models ingest, as exemplified by radiotranscriptomic mapping of perivascular fat \cite{oikonomou2019novel}. Two consequences follow for model design. First, molecular measurements are spatially sparse relative to radiomics features: omics are available only at sampled regions, not across the entire imaged field of view, so fusion architectures must tolerate this asymmetry \cite{zhou2025_multimodal_cad_review}. Second, the radiomics-guided sampling process itself introduces selection bias---tissue is preferentially obtained from regions that are abnormal on imaging---which must be acknowledged when interpreting associations between radiomics features and molecular profiles \cite{pentimalli2025challenges}.

\subsection{Early, intermediate, and late fusion}

\begin{table*}[!tp]
\centering
\sffamily\fontsize{7.5bp}{9.45bp}\selectfont
\setlength{\tabcolsep}{1.5pt}
\renewcommand{\arraystretch}{1}
\caption{Practical comparison of traditional unimodal/low-dimensional modelling versus multimodal fusion strategies for imaging-anchored multi-omics.}
\label{tab:fusion_comparison}
\begin{tabular}{@{}>{\raggedright\arraybackslash}p{\dimexpr 0.23344\textwidth - 2.250pt\relax}>{\raggedright\arraybackslash}p{\dimexpr 0.28782\textwidth - 2.250pt\relax}>{\raggedright\arraybackslash}p{\dimexpr 0.26044\textwidth - 2.250pt\relax}>{\raggedright\arraybackslash}p{\dimexpr 0.21831\textwidth - 2.250pt\relax}@{}}
\toprule
\textbf{Approach} & \textbf{Strengths} & \textbf{Common failure modes} & \textbf{When it is a good default} \\
\midrule
Classical regression/survival models with hand-crafted radiomics summaries &
High interpretability; strong calibration; stable in small samples &
Limited ability to capture spatial heterogeneity and cross-modal interactions; features may miss subtle biology &
Early translational work; limited paired data; when interpretability and calibration dominate \\[1.5pt]
Early fusion (feature concatenation) &
Simple; uses standard ML toolkits; often competitive baselines &
Sensitive to missing modalities; scaling/normalization issues; collinearity, and overfitting with high-dimensional omics &
Moderate sample sizes with mostly complete modalities; when a strong baseline is needed quickly \\[1.5pt]
Intermediate fusion (joint latent space: DCCA, multimodal VAE, and HEALNet) &
Captures non-linear cross-modal interactions; supports ``virtual omics'' and mechanistic exploration; handles missing modalities via product-of-experts &
Latent space collapse; modality imbalance; requires careful regularization and validation &
When paired subsets exist but are not large; when cross-modal generation/interpretation is desired \\[1.5pt]
Late fusion (ensembles/stacking) &
Robust to missing modalities; reuses validated unimodal models; simplifies regulatory adoption &
Cannot learn feature-level synergy; may underperform when cross-modal interactions are important; forfeits opportunity to learn from costly paired data &
When missingness is high; when unimodal models are already mature and well-validated; as deployment fallback \\[1.5pt]
Graph/contrastive and transformer-based fusion &
Naturally incorporates spatial and biological priors; can scale with pretraining; captures neighbourhood relationships lost by flat vectors &
Data-hungry; computationally expensive; prone to spurious shortcuts without strong evaluation &
When large weakly labelled data exist for pretraining, or strong spatial/biological graphs are available \\[1.5pt]
\bottomrule
\end{tabular}
\end{table*}

\textbf{Early fusion} concatenates raw or lightly processed features from all modalities into a single vector per sample before any learning occurs. In an imaging--omics cardiovascular study, this means appending CMR-derived radiomics descriptors (e.g. LGE texture, wall thickness, and strain values) to bulk gene-expression principal components or pathway activity scores, then feeding the combined vector into a standard learner such as gradient boosting, a multilayer perceptron, or a Cox survival model \cite{bhattacharya2022multi}. Early fusion is conceptually the simplest strategy and often serves as a competitive baseline. However, it assumes that all features share a comparable scale and noise structure, which is rarely the case when combining tens of radiomics descriptors with thousands of gene-expression measurements. It is also fragile to missing data: patients without one modality must be imputed or excluded entirely, which can discard a substantial fraction of the cohort given that tissue for omics is available only in selected cases~\cite{baltrusaitis2019_multimodal}.

\textbf{Intermediate fusion} first learns modality-specific embeddings, then combines them in a shared latent space---and it is this shared space that makes the strategy particularly suited to imaging--omics integration. The core idea is that radiomics features and molecular data describe the same underlying biology---tissue composition, cell-state distributions, pathway activity---but through different measurement processes; an intermediate latent space is designed to capture the biological variation that is shared across modalities while retaining private variation unique to each. Concretely, modality-specific encoders compress a CMR cine sequence into a vector summarizing ventricular geometry and wall motion, and compress a bulk gene-expression profile into a vector summarizing pathway activation patterns. These vectors are then projected into a common embedding where, for example, a fibrotic imaging phenotype and an activated-fibroblast transcriptomic programme should map to nearby points. Deep canonical correlation analysis (DCCA) achieves this by learning non-linear projections that maximize cross-modal correlation while reconstruction losses preserve modality-specific detail \cite{andrew2013_dcca,wang2015_dccae}. Multimodal VAEs with product-of-experts or mixture-of-experts inference extend this to a probabilistic setting: the encoder for each modality proposes a posterior distribution over the latent space, and these posteriors are combined via their product or mixture, so the joint latent captures information that is consistent across whichever modalities are present \cite{wu2018_multimodal_vae,joshi2022generalized}. This formulation offers three practical advantages for imaging--omics cohorts. First, the shared latent explicitly represents which aspects of the imaging phenotype are explained by measured molecular layers---and which are not---enabling mechanistic interrogation via latent traversals or factor-level association tests. Second, cross-modal generation becomes possible: omics embeddings can be decoded into predicted imaging features and vice versa, supporting ``virtual transcriptomics'' from imaging inputs. Third, the product-of-experts formulation naturally handles missing modalities, because the joint posterior remains well-defined even when only one encoder contributes, allowing patients with imaging alone or omics alone to regularize the shared space without requiring complete pairing. Recent hybrid architectures push intermediate fusion further: HEALNet uses iterative cross-attention between a shared latent bottleneck and modality-specific representations of histopathology and multi-omic data, achieving state-of-the-art survival prediction on four TCGA cancer datasets while gracefully skipping missing modalities at inference \cite{hemker2024healnet}, and the MODES framework decouples shared and modality-specific latent components from paired ECG and cardiac MRI in the UK Biobank, enabling inference of missing MRI phenotypes from ECG alone \cite{tonekaboni2025representation}.

\textbf{Late fusion} trains separate, fully independent models for each modality and combines their predictions at the decision level using averaging, voting, or a meta-learner.

\textbf{Late fusion} is sometimes viewed as wasteful for imaging--omics studies, where paired data are expensive to generate and training independent models forfeits the opportunity to learn cross-modal structure from those costly pairings. However, the strategy remains practically important for two reasons. First, it is robust to missingness: predictions can be computed from whichever modalities are available, so a patient with only CMR and no biopsy still receives a calibrated risk estimate from the radiomics arm, while the transcriptomic arm contributes when tissue is obtained---a realistic deployment scenario in which imaging is routine but omics are reserved for selected cases. Second, late fusion reuses validated unimodal models: if a radiomics-only model has already been externally validated for a fibrosis endpoint and an omics-only model has been validated for pathway-level risk scoring, their combined output inherits the credibility of each component, which simplifies regulatory and clinical adoption. Its main limitation is that it cannot capture feature-level interactions---for instance, a specific gene-expression pattern that is prognostically relevant only in the presence of a particular LGE distribution will be invisible to models that never see both modalities jointly. For imaging--omics integration, late fusion is, therefore, best understood as a robust baseline and deployment-stage fallback: it ensures that expensive molecular data are never wasted, while intermediate or graph-based fusion layers can be trained on the smaller fully paired subset to capture the cross-modal interactions that late fusion misses.

In practice, hybrid designs are common. Modality-specific encoders are often pretrained and fine-tuned on unimodal tasks, supporting late fusion, while intermediate fusion layers are trained on subsets with paired data to capture cross-modal structure. This modularity facilitates reuse and improves robustness: if a radiomics encoder is updated, the fusion and calibration layers can be retrained without discarding the transcriptomics model.

\subsection{Comparing fusion strategies with traditional approaches}

Multimodal fusion should be understood as an extension of long-standing clinical modelling approaches rather than a wholesale replacement. Classical approaches in cardiovascular risk prediction often combine a small number of radiomics-derived summary variables with demographics and labs in regression or Cox models. Fusion models generalize this by (i) allowing radiomics and omics to be represented at higher dimensionality and multiple scales, and (ii) learning non-linear interactions between modalities. Table~\ref{tab:fusion_comparison} summarizes trade-offs that are most relevant in practice: interpretability, sample size requirements, robustness to missingness, and suitability for mechanistic inference.

\subsection{Graph-based fusion and contrastive cross-modal learning}

The fusion strategies described above---early, intermediate, and late---treat each sample as a flat feature vector or a fixed-size embedding. This representation discards two types of structure that are central to imaging--omics integration in CVD. First, spatial structure: radiomics features are indexed by anatomical location (AHA segments, plaque regions, infarct zones), and spatial transcriptomic spots are connected by tissue adjacency; collapsing these into unordered vectors loses the neighbourhood relationships that determine, for example, whether a fibrotic border zone abuts viable myocardium or necrotic core. Second, biological structure: genes do not act independently but are organized into pathways, protein--protein interaction networks and regulatory modules whose topology encodes prior knowledge about coordinated function. Graph-based and contrastive fusion methods address both limitations by explicitly encoding spatial and biological relationships as edges in a graph, enabling the model to propagate and integrate information along these relationships rather than treating all features as exchangeable.

In feature graphs, nodes represent radiomics-defined imaging regions, genes, or pathways, and edges encode known relationships---such as anatomical adjacency in the heart (e.g. neighbouring AHA segments sharing a border zone) or membership in the same biological pathway (e.g. TGF-$\beta$ signalling linking fibroblast-activation genes). GNNs propagate information along these edges through iterative message passing, allowing imaging nodes to aggregate molecular signals from connected pathway nodes and molecular nodes to incorporate structural context from imaging. A graph might connect LGE-positive segments to fibrosis-related pathways enriched in bulk or spatial transcriptomics, enabling the GNN to learn which molecular programmes best explain regional fibrosis patterns. Zheng et al. \cite{zheng2024graph} demonstrated this principle by representing whole-slide pathology images as graphs whose node embeddings are fused with gene-expression signatures via cross-attention, producing spatially resolved survival activation maps that localize which tissue regions and which pathways jointly drive prognosis in non-small cell lung cancer. A comprehensive review of GNN architectures for single-cell and spatial omics---covering graph convolution, graph attention and variational graph autoencoders across more than 100 published models---confirms that graph-based encoders consistently outperform non-spatial baselines for tasks including spatial domain identification, cell-type deconvolution, and multi-omics integration \cite{li2025graph}.

In patient graphs, nodes represent patients and edges reflect similarity in radiomics features, molecular profiles, or both. Message passing across such graphs can regularize predictions, particularly in small cohorts where individual measurements are noisy \cite{parisot2018_gnn_brain}. For imaging--spatial transcriptomics integration, spatial spots can serve as nodes connected by tissue adjacency and embedded with both expression and local radiomics features, enabling graph-based segmentation of molecularly distinct tissue niches such as infarct border zones or vulnerable plaque shoulders \cite{kuppe2022_spatial_mi,campos2025_spatial_plaque}.

Contrastive multimodal learning complements graph-based fusion by providing an explicit training objective: pull together paired samples---a radiomics feature vector and its corresponding molecular profile from the same tissue region---and push apart unpaired ones, so that the learned embedding space reflects true biological correspondence rather than spurious batch or modality correlations. CLIP-style objectives have been applied to imaging--text pairs in radiology \cite{radford2021_clip,zhang2022_convirt}, and analogous objectives can be used for radiomics--omics pairing. Cross-modal autoencoders trained on paired electrocardiography, CMR, and clinical variables at biobank scale have demonstrated how a shared latent space can simultaneously support imaging-based prediction, genotype association, and counterfactual simulations \cite{crossmodal_cardiac_ae_2023}. For radiomics--spatial transcriptomics integration, contrastive objectives can align region-level radiomics embeddings with spot-level molecular embeddings, such that nearby points in latent space correspond to tissue regions that are both visually and molecularly similar. Graph contrastive learning extends this idea by treating different perturbations of the same spatial or patient graph as positives, which encourages robustness to missing spots or noisy gene counts---a property particularly valuable given the sparsity and technical noise inherent in current cardiovascular spatial transcriptomic datasets.

\subsection{Cross-modal latent spaces, missing data, and sample size}

Learning a cross-modal latent space provides a common coordinate system for downstream tasks such as clustering, subtyping, genetic association, and simulation. Multimodal VAEs, DCCA, and cross-modal autoencoders are typical tools. For imaging--omics applications in CVD, three design challenges are particularly important and interact with one another.

First, handling of missing modalities is critical. In practice, imaging is available for nearly all patients, bulk RNA-seq or proteomics may be available for a subset with tissue or blood samples, and spatial transcriptomics for a further subset where fresh-frozen sections were collected---producing a pyramid of decreasing sample size as modality depth increases. Fully paired imaging--bulk--spatial data are therefore rare. Architectures that can ingest any subset of modalities---through modality-specific encoders with masking, product-of-experts or mixture-of-experts combinations---are essential \cite{wu2018_multimodal_vae,joshi2022generalized}. These models can be trained on all available data: fully paired samples anchor the joint space, while unimodal samples regularize each encoder and reduce the risk of overfitting to small paired subsets. The PORPOISE framework demonstrated this principle at scale in oncology by fusing whole-slide histopathology images with multi-omic profiles across 14 cancer types, tolerating variable molecular availability through gated multimodal fusion and yielding interpretable prognostic markers \cite{chen2022pan}. An analogous design for cardiovascular imaging--omics would pair a CMR or echocardiographic encoder with molecular encoders and train the fusion layer on the small subset of patients with both imaging and tissue, while the imaging encoder is regularized by the much larger imaging-only cohort---precisely the strategy used by the cross-modal cardiovascular autoencoder of Radhakrishnan et al.~\cite{crossmodal_cardiac_ae_2023}, which leveraged tens of thousands of ECG--CMR pairs to stabilize a latent space that also supports genotype association.

Second, realistic expectations regarding sample size are required. High-capacity deep fusion models with millions of parameters are difficult to train on cohorts with tens of paired cases; simpler factor models, linear DCCA, or biologically constrained architectures may perform better in that regime. When the primary goal is mechanistic discovery rather than clinical prediction, modest sample sizes can still be informative if strong biological priors---pathway annotations, protein--protein interaction networks, or cell-type marker lists---are encoded in the model, for example, through pathway-level aggregation in the molecular encoder or graph-structured gene modules in the fusion layer. For patient-level prediction tasks, however, thousands of paired samples may be required to achieve stable performance and support stratified evaluation across demographic and clinical subgroups.

Third, efficient use of partially paired data is, therefore, a key strategy. Large imaging-only biobanks such as the UK Biobank ($\sim$100,000 CMR studies) and EchoNet (>10,000 echocardiographic videos) can pretrain imaging encoders, while smaller imaging--omics subsets supply the cross-modal signal needed to anchor the joint latent space. Semi-supervised designs that mix paired and unpaired data---through modality dropout during training, pseudo-labelling of unpaired samples using cross-modal predictions, or contrastive objectives that align modalities when both are present and regularize individual encoders when only one is available---extend the effective training set without requiring additional costly tissue profiling. This hierarchical data strategy, using large unimodal data for encoder pretraining and small paired data for fusion fine-tuning, is likely to be the default architecture for cardiovascular imaging--omics until the cost of spatial transcriptomics and single-cell profiling decreases substantially.

\begin{table*}[!tp]
\centering
\sffamily\fontsize{7.5bp}{9.45bp}\selectfont
\setlength{\tabcolsep}{1.5pt}
\renewcommand{\arraystretch}{1}
\caption{Evaluation metrics and validation designs commonly used (or recommended) in imaging-anchored multiomics.}
\label{tab:evaluation_metrics}
\begin{tabular}{@{}>{\raggedright\arraybackslash}p{\dimexpr 0.17733\textwidth - 2.250pt\relax}>{\raggedright\arraybackslash}p{\dimexpr 0.23171\textwidth - 2.250pt\relax}>{\raggedright\arraybackslash}p{\dimexpr 0.28591\textwidth - 2.250pt\relax}>{\raggedright\arraybackslash}p{\dimexpr 0.30506\textwidth - 2.250pt\relax}@{}}
\toprule
\textbf{Task} & \textbf{Example outputs} & \textbf{Useful metrics} & \textbf{Validation emphasis} \\
\midrule
Clinical prediction/risk modelling &
Binary outcome, time-to-event risk, treatment response &
AUC/time-dependent AUC, C-index, Brier score; calibration slope/curve; decision-curve net benefit \cite{vickers2006_decision_curve} &
External validation; leave-one-site-out; calibration and subgroup performance; reporting per TRIPOD+AI/PROBAST+AI \cite{moons2025probast} \\[1.5pt]
Radiogenomic association &
Imaging feature $\leftrightarrow$ gene/pathway score &
Effect sizes, false discovery rate control; pathway enrichment; replication rate &
Confounder adjustment; permutation tests; replication in independent cohorts \\[1.5pt]
Cross-modal alignment/retrieval &
Paired embeddings; matched region--spot correspondences &
Cross-modal retrieval (Recall@K), embedding correlation/CCA correlation, alignment error &
Robustness to missing modalities; negative controls (shuffled pairing); sensitivity to batch \\[1.5pt]
Spatial domain discovery &
Spot clusters/spatial domains &
Adjusted Rand index (if labels exist), silhouette score; spatial autocorrelation statistics &
Stability under downsampling; cross-section reproducibility; biological plausibility via marker genes \\[1.5pt]
Virtual transcriptomics &
Predicted gene/module expression maps &
Pearson/Spearman correlation, $R^2$, RMSE; spatial pattern similarity &
Hold-out donors/centres; robustness across staining/scanner protocols; uncertainty estimates \\[1.5pt]
\bottomrule
\end{tabular}
\end{table*}

\subsection{Evaluation: metrics, validation, and benchmarking}

A recurrent pitfall in multimodal bioinformatics is evaluating only the downstream prediction AUC and ignoring whether a model is calibrated, robust to distribution shift, or biologically plausible. Multimodal histology--genomic models in oncology have been shown to overestimate performance when site-of-origin batch effects are not controlled through site-preserved cross-validation \cite{howard2023multimodal}, a risk that is equally acute in cardiovascular settings where specific imaging centres also provide the only tissue samples. For imaging-anchored multiomics, evaluation should therefore be task-aligned: mechanistic models are judged by different criteria than risk models, and both require specific safeguards against confounding. Table~\ref{tab:evaluation_metrics} summarizes practical metrics and recommended validation designs across five common task types.

Beyond these task-specific metrics, three overarching evaluation principles apply to imaging--omics models. First, calibration and clinical utility should accompany discrimination: a model that achieves high AUC but assigns risk probabilities that deviate systematically from observed event rates offers limited bedside value. Decision-curve analysis quantifies whether a model provides net benefit at clinically relevant thresholds \cite{vickers2006_decision_curve}, and calibration plots should be reported alongside discrimination metrics. Second, negative controls are essential: shuffling the pairing between imaging and molecular data, or permuting spatial coordinates of transcriptomic spots, should abolish any true cross-modal signal; if the model retains apparent performance under permuted pairings, the learned association is likely confounded by batch or population structure. Third, subgroup and fairness evaluation should be standard practice: performance should be stratified by sex, ancestry, disease severity and, where possible, imaging centre, to ensure that the model does not preferentially benefit well-represented subpopulations while failing for others. Emerging AI-specific reporting guidance, including TRIPOD+AI for prediction models \cite{tripod_ai_2024} and PROBAST+AI for risk-of-bias assessment \cite{moons2025probast}, provides structured frameworks for transparent documentation of these aspects.

\subsection{Reproducibility and software ecosystem}

Reproducible multimodal fusion requires attention to data preprocessing, model specification, evaluation, and software. Preprocessing pipelines for imaging (reconstruction, segmentation, and normalization), bulk, and single-cell data (quality control, normalization, and batch correction) and spatial data (registration, spot filtering, and deconvolution) can strongly influence downstream results, and each choice introduces variation that is rarely quantified systematically. These steps should be documented, version-controlled and, where possible, implemented in reusable workflows with containerized environments and fixed random seeds.

The software ecosystem for multimodal fusion is growing but remains fragmented. General-purpose toolkits such as scvi-tools, MOFA+ and Seurat/Signac support multiomics and spatial analyses \cite{gayoso2022_scvi_tools,argelaguet2020_mofa_plus}; spatial-omics libraries such as Squidpy \cite{palla2022_squidpy} and Giotto \cite{dries2021_giotto} provide graph construction, spatial statistics, and cell--cell interaction analysis. Imaging frameworks such as MONAI \cite{monai2022} and TorchIO \cite{perezgarcia2021_torchio} handle medical image processing and model training; and open-source multimodal fusion codebases from the computational pathology community---including PORPOISE \cite{chen2022pan} and HEALNet \cite{hemker2024healnet}---provide reference implementations for histology--omics integration that can be adapted to cardiac tissue. For imaging--omics integration, no single library yet covers the full pipeline from imaging feature extraction through molecular encoding to cross-modal fusion and evaluation; most teams assemble bespoke pipelines by combining these components with custom code. Designing modular, open-source implementations that separate encoders, fusion layers, and evaluation modules---and that ship with pretrained weights where licensing permits---will be important to make imaging-anchored multiomics analyses transparent and reusable across cardiovascular centres.

Representative frameworks and exemplar studies that instantiate these integration strategies in cardiovascular settings are summarized in Table~\ref{tab:imaging_omics_frameworks}.

\begin{table*}[!tp]
\centering
\sffamily\fontsize{7.5bp}{9.45bp}\selectfont
\setlength{\tabcolsep}{1.5pt}
\renewcommand{\arraystretch}{1}
\caption{Representative frameworks and studies integrating cardiovascular imaging with molecular or other high-dimensional omic data, including general-purpose frameworks that have been applied or are directly applicable to imaging-anchored multiomics and cardiovascular studies that exemplify concrete imaging--omics applications in cardiology.}
\label{tab:imaging_omics_frameworks}
\begin{tabular}{@{}>{\raggedright\arraybackslash}p{\dimexpr 0.15760\textwidth - 2.400pt\relax}>{\raggedright\arraybackslash}p{\dimexpr 0.20069\textwidth - 2.400pt\relax}>{\raggedright\arraybackslash}p{\dimexpr 0.23899\textwidth - 2.400pt\relax}>{\raggedright\arraybackslash}p{\dimexpr 0.21160\textwidth - 2.400pt\relax}>{\raggedright\arraybackslash}p{\dimexpr 0.19111\textwidth - 2.400pt\relax}@{}}
\toprule
\textbf{Framework/study} & \textbf{Modalities integrated} & \textbf{Integration strategy} & \textbf{Cardiovascular relevance - example use} & \textbf{Key contribution} \\
\midrule
\multicolumn{5}{@{}l}{\textit{General-purpose multimodal frameworks applicable to imaging--omics integration}} \\[2pt]
MOFA+, scVI/totalVI, and scvi-tools \cite{lopez2018_scvi,gayoso2022_scvi_tools,argelaguet2020_mofa_plus} &
Bulk and single-cell transcriptomics, epigenomics, proteomics; imaging-derived features can be included as an additional view &
Probabilistic factor models (MOFA+) and hierarchical VAEs (scVI) that learn joint low-dimensional representations while explicitly modelling batch effects and missing data &
Provide molecular latent spaces---shared factors or scVI latent variables---that can be aligned with cardiac imaging embeddings for radiogenomic and spatial integration &
Established ecosystem with documented APIs; standard entry point for cardiovascular multiomics before fusion with imaging \\[1.5pt]
Multimodal VAEs \cite{wu2018_multimodal_vae,joshi2022generalized} &
Heterogeneous combinations of imaging, omics, and clinical data &
Product-of-experts or mixture-of-experts aggregation of modality-specific encoder posteriors into a shared latent space; joint posterior remains well-defined when modalities are missing &
Prototype framework for learning shared imaging--omics latent spaces; supports ``virtual omics'' generation from imaging inputs alone &
Handle missing modalities by design; probabilistic formulation enables uncertainty quantification and cross-modal simulation \\[1.5pt]
PORPOISE \cite{chen2022pan} &
Whole-slide H\&E histopathology and multi-omic profiles (RNA-seq, CNV, mutations) across 14 TCGA cancer types &
Attention-based multiple instance learning (MIL) for whole-slide images (WSIs) fused with a self-normalizing network for omics via Kronecker-product gated fusion; weakly supervised with survival labels &
Architecturally transferable to cardiac histology--omics; interpretable morpho-molecular prognostic markers at pan-cancer scale &
First large-scale interpretable histology--genomic fusion; open-source code and interactive atlas; highlighted batch-effect risks in site-unaware splits \cite{howard2023multimodal} \\[1.5pt]
HEALNet \cite{hemker2024healnet} &
Whole-slide histopathology and multi-omic data (four TCGA cancer datasets) &
Iterative cross-attention between a shared latent bottleneck and modality-specific token streams; processes raw inputs rather than precomputed embeddings &
Directly applicable to cardiac tissue sections with matched omics; handles missing modalities at both training and inference &
State-of-the-art survival prediction among end-to-end fusion models; open-source; modular design facilitates addition of imaging modalities \\[1.5pt]
\addlinespace[3pt]
\multicolumn{5}{@{}l}{\textit{Cardiovascular imaging--omics studies}} \\[2pt]
Cross-modal autoencoder for cardiovascular state \cite{crossmodal_cardiac_ae_2023} &
Twelve-lead ECG, CMR cine images and clinical traits from the UK Biobank &
Cross-modal autoencoder reconstructing each modality from a shared latent vector; supports imputation of missing modalities and genotype association in latent space &
Phenotype prediction, unsupervised GWAS, and CMR imputation from ECG at population scale (n > 40,000) &
Scalable cross-modal representation learning linking low-cost ECG signals to rich CMR phenotypes and genetic architecture \\[1.5pt]
MODES \cite{tonekaboni2025representation} &
Twelve-lead ECG and CMR from the UK Biobank &
Representation fusion framework decoupling shared and modality-specific latent components via orthogonality constraints &
Inference of missing MRI phenotypes from ECG alone; shared latent captures interpretable cardiac structure--function relationships &
Explicit shared/private decomposition improves cross-modal prediction and clinical interpretability in cardiovascular data \\[1.5pt]
Radiotranscriptomic perivascular fat signature \cite{oikonomou2018_pv_fai_risk,oikonomou2019novel} &
Coronary CTA radiomics of perivascular fat and microarray gene-expression from perivascular adipose biopsies &
Machine-learning mapping of CT radiomic features onto tissue gene-expression modules to derive a radiotranscriptomic risk score &
Prediction of cardiac mortality and MI beyond standard CT and clinical risk scores in coronary CTA cohorts &
Proof-of-concept that CT radiomics can serve as a non-invasive surrogate for local vascular inflammatory and fibrotic gene-expression states \\[1.5pt]
Spatial multi-omic map of human myocardial infarction \cite{kuppe2022_spatial_mi} &
snRNA-seq, snATAC-seq, Visium spatial transcriptomics, and histology of infarcted and control myocardium &
Joint embedding of single-cell and spatial omics to annotate infarct core, border and remote zones with dominant cell states and regulatory programmes &
Mechanistic dissection of human MI remodelling across regions and time points; tissue blocks sampled from imaging-defined zones &
Tissue-scale molecular atlas co-registrable with imaging markers of infarct size, scar distribution and remodelling \\[1.5pt]
Coronary plaque spatial transcriptomics linked to imaging \cite{campos2025_spatial_plaque,sun2023spatial,gastanadui2024_coronary_spatial_plaque} &
Spatial transcriptomics of human coronary plaques with matched histology; invasive or CT imaging in selected cohorts &
Spatial co-localization of gene-expression modules and immune niches with plaque morphology and CT features such as low-attenuation plaque or thin-cap fibroatheroma &
Characterization of high-risk plaque phenotypes and local immune microenvironments in coronary artery disease &
Template for connecting coronary imaging markers of vulnerability with spatially resolved arterial wall gene-expression and immune niches \\[1.5pt]
\bottomrule
\end{tabular}
\end{table*}

\section{Integrating imaging with transcriptomics and spatial transcriptomics}

\subsection{Radiogenomic modelling}

Radiogenomics links imaging features with molecular measurements to identify associations between non-invasive phenotypes and underlying biology. In oncology, radiogenomic analyses have associated radiomic signatures extracted from CT and MRI with tumour gene-expression programmes, mutation status and pathway activity, establishing a precedent for non-invasive molecular characterization from imaging \cite{he2024radiogenomics,guo2025images}. Analogous ideas are now emerging in cardiovascular medicine, where the aim is to connect imaging-derived features---regional LGE burden, wall thickness, strain patterns, plaque attenuation profiles---with transcriptomic, proteomic, or pathway-level molecular readouts \cite{lin2025machine}. A systematic review of cardiac CT- and CMR-based radiogenomics found that, while radiomic features showed good diagnostic accuracy for plaque and myocardial characterization, very few studies had yet exploited full radiogenomic integration in cardiology, highlighting a substantial gap that imaging-anchored multiomics aims to address \cite{infante2021radiogenomics}.

\begin{figure*}[!t]
\centering
\includegraphics[width=\textwidth]{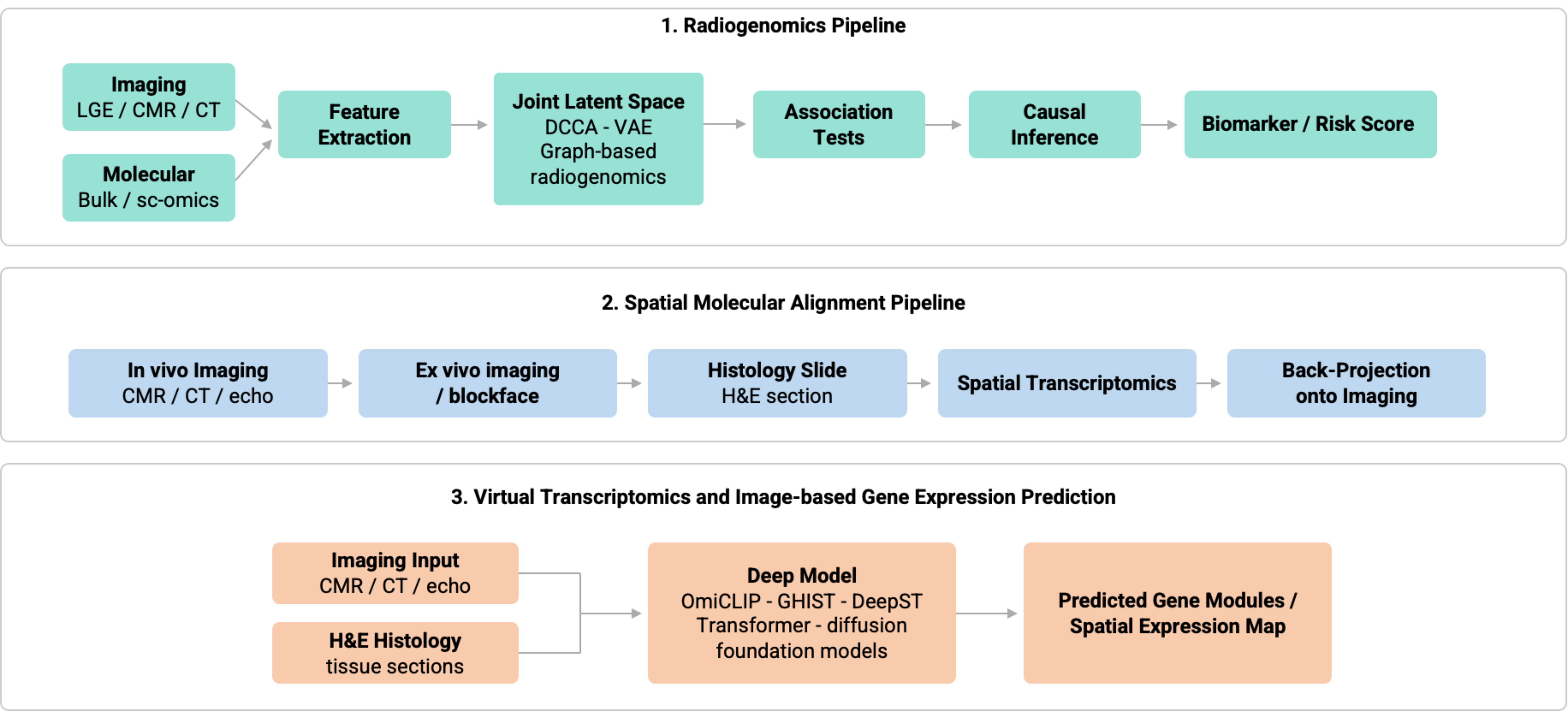}
\caption{Archetypal pipelines for imaging--omics integration. Radiogenomics workflows link imaging-derived features to bulk or single-cell molecular profiles via feature extraction, joint latent modelling, and association or causal inference; spatial molecular alignment pipelines register in vivo imaging, ex vivo imaging, histology and spatial transcriptomics into a common coordinate system; virtual transcriptomics approaches use deep models to predict molecular modules or spatial expression maps directly from imaging inputs.}
\label{fig:integration_pipelines}
\end{figure*}

At its simplest, radiogenomic analysis treats imaging-derived features as predictors and gene-expression or pathway scores as outcomes, or conversely uses molecular profiles to predict imaging phenotypes. Regression, correlation, and enrichment analyses can identify genes and pathways whose expression covaries with imaging phenotypes such as myocardial fibrosis burden or plaque vulnerability. The most developed cardiovascular example to date is the radiotranscriptomic mapping of perivascular fat by Oikonomou et al.~\cite{oikonomou2018_pv_fai_risk,oikonomou2019novel}, who trained machine-learning models to map coronary CTA radiomic features of perivascular adipose tissue onto tissue-level gene-expression modules obtained from matched biopsies, deriving a radiotranscriptomic signature that improved cardiac risk prediction beyond standard CT and clinical scores. This study illustrates the core radiogenomic workflow: imaging features define a non-invasive phenotype, paired tissue profiling provides the molecular ground truth, and a learned mapping enables molecular inference from imaging alone in subsequent patients without biopsies. The major radiogenomic, spatial-alignment, and virtual-transcriptomics pipelines covered in this section are shown in Fig.~\ref{fig:integration_pipelines}.

More sophisticated models place imaging and transcriptomics into a joint latent space. DCCA and deep canonical autoencoders learn non-linear transformations that maximize cross-modal correlation \cite{andrew2013_dcca,wang2015_dccae}. Multimodal VAEs with product-of-experts inference can generate imaging-like representations from transcriptomic inputs and infer transcriptomic profiles from imaging, providing a generative linking model \cite{wu2018_multimodal_vae,joshi2022generalized}. These approaches are particularly useful when imaging and omics are not perfectly aligned in time---a common situation in cardiology, where late-stage transcriptomic profiles from explanted or surgically sampled hearts must be related back to earlier imaging phenotypes captured months or years prior. The cross-modal autoencoder of Radhakrishnan et al. \cite{crossmodal_cardiac_ae_2023} demonstrated this temporal bridging at population scale by learning a shared latent space across ECG, CMR, and clinical variables in the UK Biobank, enabling phenotype prediction and genotype association in the latent space despite non-simultaneous data acquisition.

Graph-based radiogenomics introduces biological priors by representing genes, pathways, and imaging features as nodes in a graph, with edges encoding known relationships such as protein--protein interactions, pathway co-membership, or anatomical adjacency. GNNs trained on such graphs can propagate information between molecular and imaging nodes, learning which molecular programmes best explain regional imaging patterns. To illustrate, consider hypertrophic cardiomyopathy (HCM), where the individual components for such a graph already exist in the literature: sarcomeric gene variants in MYBPC3 and MYH7 have been associated with distinct patterns of LGE extent and myocardial fibrosis on CMR \cite{kramer2015_hcmr,ho2018genotype}, and single-cell transcriptomics of HCM myectomy tissue has revealed cell-type-specific gene networks---including stress, hypertrophic, and fibrotic programmes in cardiomyocytes and fibroblasts---that drive the remodelling phenotype \cite{wehrens2022single}. A graph-based radiogenomic model could tie together genotype nodes (sarcomeric variant status), imaging nodes (regional LGE distribution, wall thickness, and strain) and transcriptomic nodes (fibroblast activation and cardiomyocyte stress programmes) to explain why specific genotypes produce characteristic imaging patterns and to distinguish pathogenic from benign variants---although such an end-to-end integrative model has not yet been demonstrated in HCM and remains a methodological opportunity.

Causal frameworks such as Mendelian randomization extend radiogenomics beyond association. Genetic variants associated with biomarkers or gene-expression modules can be tested as instruments for imaging traits, probing whether variation in a molecular mediator causally influences an imaging phenotype. Conceptually, this allows chains such as genotype $\rightarrow$ transcriptomic module $\rightarrow$ imaging phenotype $\rightarrow$ outcome to be decomposed and quantified~\cite{ahlberg2024_cmr_gwas}. These analyses are limited by instrument strength and pleiotropy but add a layer of mechanistic interpretation beyond purely correlational radiogenomic studies, and are increasingly feasible at scale as biobank-linked imaging and omics cohorts grow \cite{petersen2013imaging,bai2020population}. Recent work demonstrates the potential of such scale: a multitask deep learning framework trained on UK Biobank proteomics and metabolomics data predicted six common CVDs up to 15 years before clinical onset and identified disease-relevant proteins and metabolites as candidate biomarkers \cite{luo2026ai}. These developments underscore that the computational and data infrastructure for cardiovascular radiogenomics at population scale is maturing rapidly, even as paired imaging--transcriptomic cohorts remain the principal bottleneck.

\subsection{Spatial molecular alignment and tissue--image correspondence}

Linking spatial transcriptomics to in vivo imaging requires careful alignment across scales: from MRI voxels or echocardiographic pixels to histology, from histology to array capture spots and from spots to inferred cell types \cite{khan2025comprehensive}. Experimental design is crucial. In some studies, tissue blocks are sampled from imaging-identified regions such as LGE-positive and LGE-negative segments, which establishes a coarse correspondence between imaging and spatial maps \cite{kuppe2022_spatial_mi,calcagno2022_infarct_spatial}. In others, high-resolution ex vivo MRI or micro-CT is acquired on explanted hearts as an intermediate for registration, bridging the resolution gap between in vivo imaging and the histology section on which spatial transcriptomics is performed \cite{pentimalli2025challenges}.

Computationally, tissue--image alignment often proceeds in stages. First, histology slides are registered to block-face photographs or ex vivo imaging using affine and non-linear transformations. Second, the spatial transcriptomics capture grid is mapped onto the histology image using fiducial markers or deep-learning-based landmark detection algorithms that can handle nonlinear tissue deformations across modalities \cite{palla2022_squidpy,ekvall2024spatial}. Third, in vivo imaging is registered to ex vivo imaging or to 3D histology reconstructions via anatomical landmarks and deformable warping. At each step, uncertainties arise from tissue deformation, sectioning artefacts, and differences in imaging physics \cite{pentimalli2025challenges}. Representing these uncertainties explicitly, through probabilistic registration or ensembles of transformations, is important when assessing spatial correlation between imaging and molecular features. Recent work on integrating tissue morphology with spatial transcriptomics further highlights how histology foundation models can enhance alignment and spatial domain identification, although their application to cardiac tissue remains nascent \cite{chelebian2025combining}.

Once correspondence is established, imaging-derived quantities such as local LGE intensity, strain, or CT attenuation can be sampled at projected spot coordinates, producing paired vectors of imaging and expression features for each spatial location. Spatial graph encoders can then jointly embed these paired features, allowing discovery of spatial domains that are coherent in both imaging and molecular space and aiding identification of structures such as border zones, fibrotic microdomains, or vulnerable plaque shoulders \cite{kuppe2022_spatial_mi,campos2025_spatial_plaque}.

\subsection{Image-based gene-expression prediction and virtual transcriptomics}

An emerging direction is to predict molecular profiles directly from images, often termed virtual transcriptomics, or image-based gene-expression imputation. In atherosclerosis, models trained on coronary CT angiography and matched plaque gene-expression profiles have predicted expression of inflammatory and matrix-degrading genes from CT alone, effectively performing a non-invasive molecular biopsy of the plaque \cite{oikonomou2019novel}. Similar approaches in oncology predict bulk or spatial transcriptomic patterns from radiology or histology images~\cite{chen2025visual,cisternino2024self,fu2025spatial}.

For cardiac applications, myocardial or plaque gene-expression modules could be predicted from imaging features such as LGE patterns, T1 maps, or strain fields. This would enable large-scale molecular phenotyping in cohorts where only imaging is available. Challenges include the high dimensionality of gene-expression outputs, often addressed by predicting pathway scores or modules rather than individual genes, limited numbers of paired image--omics samples and domain differences between training and deployment data. A recent benchmark of 11 histology-to-expression prediction methods confirmed that H\&E-based virtual transcriptomics is feasible across tissue types, while identifying batch effects and limited cross-dataset generalizability as key remaining challenges \cite{wang2025benchmarking}. Several deep learning methods predict high-dimensional spatial transcriptomic profiles directly from histology, often using transformer or diffusion backbones trained in a semi-supervised manner on large spatial atlases \cite{heist_2025,deepst_2022}; vision--omics foundation models such as OmiCLIP further demonstrate that spatial gene expression can be predicted at Visium spot level from H\&E patches alone after training on millions of paired image--transcriptome samples across 32 organs \cite{chen2025visual}, and GHIST achieves single-cell-resolution gene-expression prediction from histology using deep learning \cite{fu2025spatial}. These developments demonstrate that high-fidelity virtual transcriptomics is technically feasible and provide architectures that can be adapted to cardiac histology, CMR, or CT once joint image--omics datasets reach adequate scale.

\subsection{Applications: subtyping, mechanism, biomarkers, and early clinical case studies}

Integrative imaging--transcriptomic pipelines support four recurring goals: disease subtyping, mechanistic insight, biomarker discovery and virtual transcriptomics, and early clinical translation.

\subsubsection{Disease subtyping}

Clustering in cross-modal latent spaces allows grouping of patients or tissue regions based on joint imaging and molecular characteristics. In heart failure, integrating echocardiographic phenotypes with transcriptomic or proteomic modules has revealed phenogroups that differ both in structure and in underlying biology; one subgroup may have predominantly inflammatory signatures, whereas another is characterized by fibrotic changes \cite{koenig2022_cardiomyopathy_rnaseq}. Spatial transcriptomics of failing human myocardium has further resolved region-specific molecular programmes within hearts that appear homogeneously dilated on imaging, revealing distinct spatial domains of fibrosis, inflammation, and metabolic remodelling \cite{lee2025_spatial_hf}. Spatially, clustering of spots based on combined imaging and expression features can identify microdomains, such as stressed border-zone myocardium, that are not evident from histology or gene expression alone \cite{kuppe2022_spatial_mi}.

\subsubsection{Mechanistic discovery}

Radiogenomic and spatial analyses can suggest mechanisms for imaging findings. Spatial transcriptomics of infarct border zones linked to LGE and strain abnormalities has revealed co-localized programmes of fibroblast activation, angiogenesis, and immune-cell recruitment \cite{kuppe2022_spatial_mi,calcagno2022_infarct_spatial}. In atherosclerotic plaques, combining CT features with spatial expression maps has implicated specific macrophage and smooth muscle cell states in regions corresponding to low-attenuation plaque and thin fibrous caps \cite{campos2025_spatial_plaque,sun2023spatial}. Such findings help explain which cell types and pathways drive particular imaging phenotypes and how they might be targeted therapeutically.

\subsubsection{Biomarker discovery}

Imaging phenotypes that predict outcomes can be mined for molecular correlates, yielding candidate biomarkers. When a specific pattern of non-ischaemic fibrosis on CMR is strongly associated with arrhythmic events, imaging--omics integration can identify gene-expression modules or cell-state proportions enriched in regions with that pattern. These molecular signatures can then be sought in blood-based assays---such as plasma proteomics, where protein-based risk scores have already demonstrated incremental prediction of cardiovascular events beyond classical factors \cite{ganz2016_protein_risk_jama}---or used to stratify patients in clinical trials. A recent multitask deep learning framework integrating proteomics and metabolomics from the UK Biobank predicted six common CVDs $\sim$15 years before onset and identified disease-relevant circulating biomarkers \cite{luo2026ai}, illustrating how multiomics-derived molecular signatures can complement imaging phenotypes for early risk stratification. Conversely, known circulating biomarkers can be traced back to spatial imaging patterns, clarifying which structural or cellular changes they reflect.

\subsubsection{Early clinical case studies}

A notable example of imaging-anchored molecular inference is the perivascular FAI, developed from coronary CTA as a surrogate for local coronary inflammation and shown to predict residual cardiovascular risk \cite{oikonomou2018_pv_fai_risk}. Subsequent work derived a radiotranscriptomic signature by mapping perivascular adipose gene-expression modules to CTA radiomic features, improving risk prediction beyond standard models in coronary CTA cohorts \cite{oikonomou2019novel}. In a complementary direction, cross-modal representation learning at biobank scale has shown that a shared latent space across ECG and CMR can enable MRI imputation and genetic association studies in the latent space \cite{crossmodal_cardiac_ae_2023}, and the MODES framework further demonstrated that explicitly decoupling shared and modality-specific latent components improves cross-modal prediction of missing MRI phenotypes from ECG alone \cite{tonekaboni2025representation}. Together, these case studies illustrate that (i) imaging can act as a proxy for local biology in specific contexts, and (ii) large-scale paired datasets can support clinically relevant cross-modal phenotyping.

\subsection{Predictive learning, risk models, and reporting considerations}

Imaging-anchored multiomics can be used for mechanistic biology, but many translational applications ultimately require predictive models: outcome prediction, treatment response, and risk stratification. Unimodal AI-based predictors have already demonstrated that routine clinical signals carry far more prognostic information than conventional interpretation extracts: deep neural networks trained on resting 12-lead ECGs predict one-year all-cause mortality with an AUC of 0.88, including among tracings read as ``normal'' by cardiologists \cite{raghunath2020prediction}; analogous models detect paroxysmal atrial fibrillation from sinus-rhythm recordings \cite{attia2019artificial}; and the AIRE platform extends AI-ECG prediction to multiple actionable cardiovascular endpoints with transnational external validation \cite{sau2024artificial}. In echocardiography, multitask deep learning now automates comprehensive interpretation across diverse pathologies, with a multi-view vision--language foundation model achieving state-of-the-art performance on 23 benchmarks of cardiac form and function \cite{vukadinovic2026echoprime}. These unimodal successes establish the ceiling that multimodal fusion must surpass to justify its additional complexity.

Multimodal and multi-omic predictive models are beginning to meet this bar. A machine-learning marker for coronary artery disease derived from electronic health records and validated in two independent biobank cohorts quantified atherosclerotic burden on a continuous scale and identified underdiagnosed individuals at elevated mortality risk \cite{forrest2023machine}. In hypertrophic cardiomyopathy, the MAARS model fused electronic health records, echocardiographic and radiology reports, and contrast-enhanced CMR through transformer-based networks to forecast lethal ventricular arrhythmia with an AUC of 0.89 internally and 0.81 externally, outperforming current guideline-based risk calculators \cite{lai2025multimodal}. At the multiomics level, integrating genomics, transcriptomics, proteomics and metabolomics into a single predictive framework has demonstrated that each omic layer contributes non-redundant information to cardiovascular risk, with the combined model significantly outperforming any single modality \cite{luo2026ai}. Collectively, these studies show that multimodal fusion adds genuine prognostic value when validated rigorously.

Four practical points are central to translating these models. First, multimodal models should be evaluated not only for discrimination but also for calibration and clinical utility. A model that achieves high AUC but assigns risk probabilities that deviate systematically from observed event rates offers limited bedside value. Calibration determines whether predicted risks are meaningful in clinical decision-making; decision--curve analysis quantifies whether a model provides net benefit at clinically relevant thresholds \cite{vickers2006_decision_curve}. External validation across centres and demographic subgroups is essential, because multimodal histology--genomic models in oncology have been shown to overestimate performance when site-of-origin batch effects are not controlled through site-preserved cross-validation \cite{howard2023multimodal}---a risk equally acute in cardiovascular settings where specific imaging centres often provide the only tissue samples. Leave-one-site-out designs and negative controls with permuted pairings should be standard practice for imaging--omics models, as discussed in the ``Bottlenecks, root causes, and failure modes'' section.

Second, validation should be stratified. Performance should be reported across clinically meaningful subgroups---by sex, ancestry, disease severity, and imaging centre---to ensure that the model does not preferentially benefit well-represented subpopulations while failing for others. The radiotranscriptomic perivascular fat signature, for example, was validated in independent coronary CTA cohorts and shown to improve risk prediction beyond standard clinical and imaging scores \cite{oikonomou2019novel}, while the cross-modal cardiovascular autoencoder was evaluated across tens of thousands of UK Biobank participants for phenotype prediction and genotype association \cite{crossmodal_cardiac_ae_2023}. These examples illustrate the scale and rigour of validation needed for clinical translation.

Third, reporting and risk-of-bias assessment for predictive models should follow emerging AI-specific guidance. TRIPOD+AI extends the TRIPOD statement to prediction models developed using machine learning \cite{tripod_ai_2024}, and PROBAST+AI adapts risk-of-bias assessment to AI-driven prediction modelling \cite{moons2025probast}. For multimodal imaging--omics models, these frameworks emphasize transparent reporting of data provenance, missingness mechanisms, preprocessing, model selection, and validation strategy. Emerging guidance for generative clinical AI further highlights the need for detailed documentation of intended use and limitations \cite{sushil2025_miclaim_gen}.

Fourth, as multimodal architectures scale from unimodal baselines through pairwise fusion to full imaging--omics integration, the taxonomy of fusion challenges---representation, alignment, translation, and co-learning \cite{baltrusaitis2019_multimodal}---provides a useful design checklist. Each added modality increases data requirements and potential failure modes; demonstrating that the multimodal model outperforms each unimodal component, not just a naïve concatenation, remains the minimum standard for claiming added value.

\subsection{Bottlenecks, root causes, and failure modes}

Sample sizes of deeply phenotyped patients with paired imaging, bulk, single-cell, and spatial data remain small---typically tens to low hundreds of patients---which limits statistical power and encourages overfitting. Cohort composition often reflects specific indications, such as transplant or surgical patients, which may restrict generalizability to broader cardiovascular populations. Variability in tissue sampling relative to imaging, such as sampling only a single septal biopsy in a heart with patchy disease, can bias associations; CMR-guided endomyocardial biopsy improves spatial targeting \cite{unterberg2017targeted} but remains technically demanding and is not standard practice. Spatial misalignment and tissue distortion from fixation, sectioning, and mounting can introduce errors in spot-to-voxel mapping that blur true correlations, a challenge well characterized in the spatial transcriptomics registration literature \cite{khan2025comprehensive,ekvall2024spatial}. These challenges are shared broadly across multimodal biomedical AI, where data heterogeneity, missingness, and privacy constraints compound modelling difficulty at every stage from representation to deployment~\cite{acosta2022multimodal}.

Many failure modes trace back to a small set of root causes; below, we pair each with practical mitigations.

\textbf{(i) Mis-specified correspondence across scales.} Even small registration errors---a few voxels in MRI, or rotational offsets in histology---can wash out true imaging--molecular relationships, particularly for spatially heterogeneous phenotypes such as patchy fibrosis or border-zone remodelling. In myocardial infarction studies, infarct core, border zone, and remote regions lie within millimetres of each other \cite{kuppe2022_spatial_mi}, making registration fidelity critical. Mitigations include explicit uncertainty modelling through ensembles of registrations, sensitivity analyses that quantify how results change under plausible warps, deep learning-based spatial landmark detection for automated tissue alignment \cite{ekvall2024spatial}, and experimental designs that sample tissue using imaging-defined targets---such as LGE-positive versus LGE-negative segments---whenever possible \cite{unterberg2017targeted}.

\textbf{(ii) Confounding and shortcut learning.} Site, scanner, and processing batch can dominate latent factors, particularly when paired sample sizes are limited. Multimodal histology--genomic deep learning models such as pan-cancer integrative frameworks \cite{chen2022pan} have been shown to overestimate performance when site-of-origin batch effects are not controlled through site-preserved cross-validation \cite{howard2023multimodal}; the same risk applies to cardiovascular cohorts where one centre contributes all paired imaging--spatial data. Imaging domain shifts arise from differences in scanners, field strength, acquisition protocols, and reconstruction algorithms; deep learning harmonization methods for structural MRI are developing but remain less mature than omics batch correction \cite{abbasi2024deep}. Mitigations include batch-aware latent models such as ComBat \cite{johnson2007_combat} and Harmony \cite{korsunsky2019_harmony} for omics, domain-adversarial training for imaging, leave-one-site-out evaluation, and negative controls with randomized pairings or permuted spatial coordinates.

\textbf{(iii) Modality imbalance and missingness.} Deep fusion models can collapse onto the strongest modality (typically imaging, which has the largest sample size and richest spatial detail) or exclude patients with missing data, biasing the cohort toward fully profiled subsets. In practice, a cohort may have CMR on thousands of patients, bulk RNA-seq on hundreds, and spatial transcriptomics on dozens; naïve training on the intersection discards most of the imaging data. Mitigations include product-of-experts or mixture-of-experts inference that naturally handles variable input subsets \cite{wu2018_multimodal_vae,joshi2022generalized}, modality dropout during training (randomly masking one modality per batch to prevent co-adaptation), explicit missingness indicators, and late-fusion baselines that combine predictions from whichever modalities are available. Leveraging large imaging-only biobanks alongside smaller imaging--omics subsets is a key design strategy \cite{crossmodal_cardiac_ae_2023}.

\textbf{(iv) Over-interpretation of correlational links.} Radiogenomic associations can be mistaken for causality. A strong correlation between an imaging feature and a gene-expression module may reflect shared confounders (e.g. disease severity driving both) rather than a mechanistic link. Mitigations include replication in independent cohorts, triangulation with genetic instruments via Mendelian randomization where feasible \cite{ahlberg2024_cmr_gwas}, experimental follow-up in model systems, and careful separation of discovery and validation datasets to prevent circular reasoning.

\textbf{(v) Reproducibility gaps.} Heterogeneous preprocessing and analysis choices---segmentation algorithms, normalization strategies, gene-filtering thresholds, spot quality-control criteria---can alter conclusions substantially. Spatial transcriptomics pipelines are particularly sensitive to the choice of deconvolution method and reference atlas \cite{pentimalli2025challenges}. Emerging foundation models that couple imaging, text and structured data across tasks \cite{moor2023_foundation_models} will require particularly rigorous auditing given their scale and opacity. Mitigations include workflow tooling with containerized environments, versioned preprocessing, public code and model cards, and adherence to reporting guidance for predictive models such as TRIPOD+AI and PROBAST+AI \cite{tripod_ai_2024,moons2025probast}. A general taxonomy of multimodal learning challenges---representation, alignment, translation, fusion, and co-learning \cite{baltrusaitis2019_multimodal}---provides a useful checklist for identifying where reproducibility investments are most needed.

These root causes are not independent: small sample sizes (the overarching constraint) amplify every other failure mode, making confounding harder to detect, missingness harder to impute, and registration errors harder to average out. Addressing them requires coordinated advances in cohort design (multi-centre, \emph{imaging-guided tissue sampling}), computational methodology (robust fusion under missingness and batch effects), and community standards (open data, shared preprocessing pipelines, and standardized evaluation protocols).

\section{Multimodal transformers and agentic systems}

\subsection{Multimodal transformers for imaging--omics}

Multimodal transformers generalize fusion by representing heterogeneous inputs as tokens processed by self-attention. Imaging patches, gene modules, spatial spots, and clinical variables can each be projected into a shared token space with modality-specific and positional encodings, then integrated via shared attention layers that learn cross-modal interactions without requiring hand-crafted feature alignment \cite{moor2023_foundation_models,acosta2022multimodal}. Masking enables training and inference with variable subsets of modalities---a patient with only CMR and bulk RNA-seq receives predictions from the same model that handles a fully profiled case with additional spatial transcriptomics---while attention weights can highlight influential imaging regions, pathways, or cell types, offering a route to post-hoc interpretability.

Surveys emphasize these architectures as a core component of emerging medical foundation models that integrate imaging, text and structured data \cite{moor2023_foundation_models,tu2024_generalist_biomed_ai,foundation_radiology_2025,sun2024_medical_mmfm}. In oncology, multimodal transformers that tokenise histology patches alongside genomic features have demonstrated improved survival prediction across cancer types \cite{chen2022pan}; the HEALNet architecture further shows that heterogeneous biomedical modalities can be fused through iterative cross-attention without requiring all modalities at test time \cite{hemker2024healnet}. In cardiology, transformer-based networks underpin both the MAARS model for arrhythmic death prediction from CMR, echocardiographic reports, and electronic health records \cite{lai2025multimodal}, and EchoPrime, a view-informed model that integrates multiple echocardiographic views for comprehensive study-level interpretation \cite{vukadinovic2026echoprime}. A vision--language foundation model for echocardiography further demonstrates how contrastive pretraining on paired echo videos and clinical text can yield transferable cardiac representations \cite{christensen2024vision}.

In parallel, single-cell and spatial foundation models pretrained on tens of millions of cells---such as scGPT \cite{cui2024scgpt} and Nicheformer \cite{nicheformer_2025}---provide transferable molecular encoders whose embeddings can serve as token inputs to a multimodal transformer alongside imaging tokens, bridging the resolution gap between voxel-level imaging and cell-level transcriptomics.

Despite these advances, multimodal transformers remain data-hungry and computationally intensive relative to typical paired cardiovascular cohort sizes. A pragmatic strategy is to pretrain unimodal transformers (imaging, single-cell) using self-supervision on large unlabelled archives, then train a lighter cross-modal attention or fusion transformer on the available paired data. This modular approach allows each encoder to benefit from the scale of its own modality while limiting the parameter count of the fusion layers to what the paired subset can support.

\subsection{Agentic and large language model-based orchestration}

Large language models (LLMs) are increasingly used to orchestrate complex analysis pipelines by calling specialized tools. In an imaging-anchored multiomics workflow, encoders, fusion models, and statistical tests can be wrapped as callable tools, and an LLM-based agent can coordinate data retrieval, model execution, result interpretation, and report drafting \cite{quer2024_llm_cardiovascular_medicine}. Such orchestration is particularly attractive for imaging--omics pipelines that span multiple software ecosystems---imaging frameworks such as MONAI \cite{monai2022}, single-cell toolkits such as scvi-tools \cite{gayoso2022_scvi_tools}, and spatial analysis libraries such as Squidpy \cite{palla2022_squidpy}---where an agent can automate the handoffs between preprocessing, encoding, fusion, and evaluation.

LLMs should not replace validated integration models: current limitations including hallucinations, brittle multi-step reasoning, and lack of calibrated uncertainty motivate constrained tool use, structured logging, and mandatory human oversight at decision points \cite{sarraju2023_llm_cardiology}. Regulatory considerations, including the European Union AI Act and guidance for regulated digital medical products, further emphasize traceability, auditability, and human oversight for clinical-facing agentic systems \cite{eu_ai_act_2024}. The MI-CLAIM-GEN checklist for generative clinical AI provides a reporting framework that is directly applicable to agentic pipelines that generate clinical interpretations~\cite{sushil2025_miclaim_gen}.

\subsection{Cross-domain methodological inspirations}

Design patterns from adjacent biomedical AI domains can inform imaging--omics integration. Dual-feature attention fusion, developed for anticancer natural product detection, demonstrates how two-sided attention across heterogeneous feature sets can capture cross-modal synergies \cite{norouzi2025dft_anpd}. Explicit part--whole anatomical priors, introduced for scene parsing, offer a template for encoding the hierarchical organization of cardiac anatomy (chambers, walls, and segments) into fusion architectures \cite{abbasi2020_partwhole_fcn}. Systematic perspectives on molecular representations and evaluation in generative modelling highlight best practices for assessing the fidelity of predicted molecular outputs \cite{dmytrenko2025_comp_drug_design_review}, which are directly relevant when virtual transcriptomics models generate gene-expression profiles from imaging inputs. The broader taxonomy of multimodal machine learning challenges---representation, alignment, translation, fusion, and co-learning \cite{baltrusaitis2019_multimodal}---provides an organizing framework that unifies these cross-domain inspirations and helps identify which design patterns are most transferable to cardiovascular imaging--omics.

\section{Practical considerations}

\subsection{Data cleaning, harmonization, and overall principles}

Robust imaging--omics integration starts with rigorous data curation. For imaging this includes quality control, consistent reconstruction, and careful segmentation or landmarking with clinical oversight; frameworks such as MONAI \cite{monai2022} and TorchIO \cite{perezgarcia2021_torchio} provide standardized pipelines for medical image preprocessing, augmentation, and patch-based sampling that improve reproducibility across sites. For bulk and single-cell data, curation includes alignment and quantification, filtering low-quality cells and genes, and appropriate normalization; toolkits such as Scanpy \cite{wolf2018scanpy} and scvi-tools \cite{gayoso2022_scvi_tools} offer end-to-end workflows with integrated quality-control metrics. Spatial transcriptomics data add further requirements: tissue coverage checks, spot-level quality control, accurate registration to paired histology, and careful choice of deconvolution method, all of which can materially alter downstream conclusions \cite{pentimalli2025challenges}.

Harmonization is critical across modalities. Omics pipelines commonly use ComBat \cite{johnson2007_combat}, Harmony \cite{korsunsky2019_harmony}, or joint latent models such as scVI \cite{lopez2018_scvi} to reduce technical variation while preserving biological signal. Imaging harmonization is less standardized but may include intensity normalization, resampling, and domain-adversarial training to reduce scanner-specific bias; deep learning-based approaches for structural MRI harmonization are developing but not yet mature \cite{abbasi2024deep}. Across all steps, adherence to Findable, Accessible, Interoperable, Reusable principles and rich metadata---acquisition protocols, preprocessing versions, quality-control metrics, and provenance records---improve reproducibility, reuse, and the ability to diagnose failures post hoc.

\subsection{Computational demands, efficiency, and scalability}

Compute constraints shape feasible designs and should be reported alongside model performance. Radiomics and regression-style radiogenomics can run on modest hardware, while 3D/4D imaging encoders, multimodal transformers, and large spatial graph models may require multi-GPU training and significant storage for intermediate features and graphs. The gap between the computational envelope of foundation models---pretrained on millions of images or cells \cite{moor2023_foundation_models,singlecell_foundation_2024}---and the resources available to most cardiovascular research groups is substantial.

Practical approaches to bridge this gap include: (i) pretraining unimodal encoders once on large archives and freezing them during fusion, so that only lightweight cross-modal layers are trained on scarce paired data; (ii) operating at clinically meaningful intermediate resolutions---American Heart Association segments or spatial domains rather than individual voxels or spots---when paired data are limited; (iii) predicting pathway or module-level scores rather than thousands of individual genes, which reduces output dimensionality and improves interpretability; and (iv) distilling large models into smaller, deployable surrogates for clinical integration. Reporting training compute and memory footprints, along with lightweight baselines, improves feasibility assessment, and reproducibility \cite{tripod_ai_2024}.

\subsection{Cross-centre variability and deployment}

Multicentre cohorts differ in demographics, disease mix, imaging hardware, and omics platforms; single-centre models often fail to generalize \cite{acosta2022multimodal}. This is not merely a technical inconvenience: in cardiovascular imaging--omics, a model trained at a single transplant centre may learn features specific to end-stage disease and advanced immunosuppression that do not transfer to community-level screening. Pipelines should incorporate site and batch covariates, domain-adaptation where appropriate, and external validation following TRIPOD+AI guidance \cite{tripod_ai_2024}. Federated learning can reduce barriers to cross-site training by keeping data local, but must handle statistical heterogeneity across sites and the operational complexity of coordinating preprocessing and model updates.

Cross-modal autoencoders trained at biobank scale have demonstrated that shared latent spaces across ECG, CMR, and clinical variables can scale to tens of thousands of participants when designed for partially paired data \cite{crossmodal_cardiac_ae_2023}, providing a template for how large imaging-only cohorts such as UK Biobank \cite{petersen2013imaging} can be leveraged alongside smaller imaging--omics subsets. Deployment additionally favours models that are calibrated, interpretable, efficient, and maintainable. In some settings, simplified surrogate scores distilled from deep models may be necessary for clinical integration, particularly when real-time inference or point-of-care use is envisaged.

\subsection{Bias, ethics, and equity}

Multimodal datasets can reflect and amplify healthcare disparities. Under-representation of women, minority ethnic groups and low-resource settings can lead to biased models or unreliable molecular interpretations---a concern that extends across all of biomedical AI \cite{acosta2022multimodal} but is particularly acute in imaging--omics, where the most deeply phenotyped cohorts tend to be drawn from well-resourced academic centres in high-income countries. Performance should be assessed across demographic and clinical subgroups, and uncertainty should be reported alongside predictions; multimodal models that appear accurate in aggregate may fail systematically for under-represented groups, as has been documented for batch-confounded histology--genomic models in oncology \cite{howard2023multimodal}.

Transparent documentation of intended use, data provenance, and limitations is essential. Emerging guidance for AI-driven prediction models---TRIPOD+AI \cite{tripod_ai_2024} and PROBAST+AI \cite{moons2025probast}---and for generative clinical AI \cite{sushil2025_miclaim_gen} provide reporting frameworks that are directly applicable to imaging-anchored multiomics. For this field, avoiding over-interpretation of correlational links and ensuring that clinical benefit generalizes across populations are core ethical requirements that should be treated as design constraints rather than afterthoughts.

\section{Future directions}

\begin{figure*}[!t]
\centering
\includegraphics[width=\textwidth]{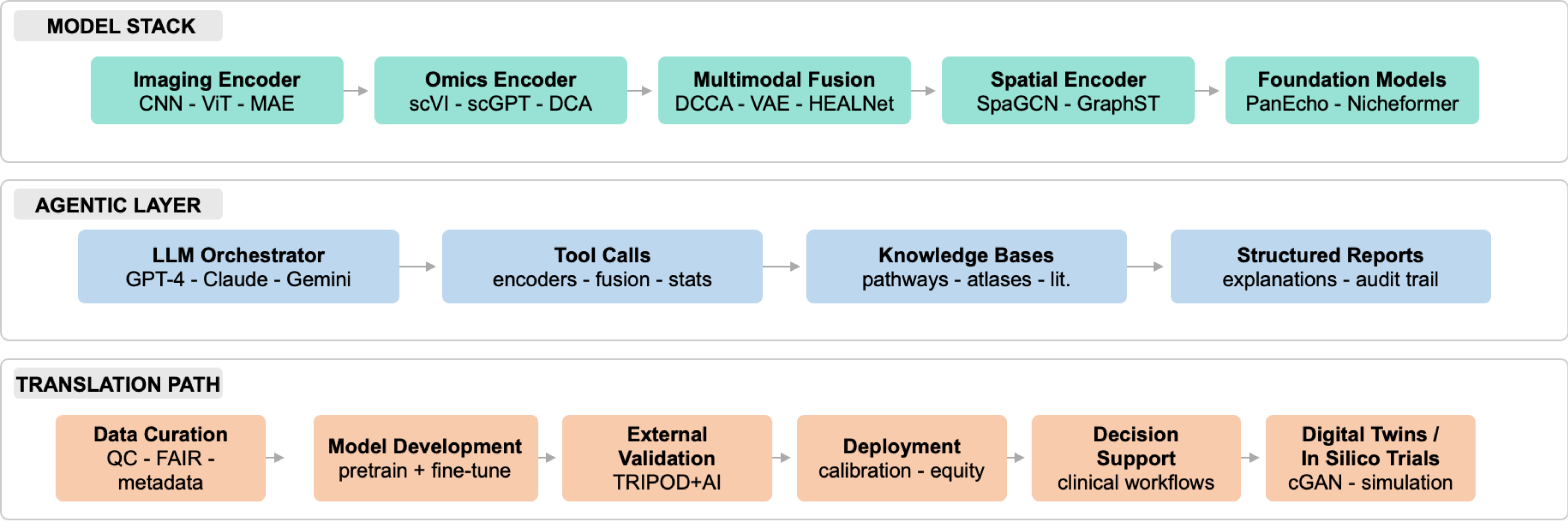}
\caption{Translational roadmap from imaging-anchored multiomics to digital twins. Top lane: model stack comprising modality-specific encoders, multimodal fusion models and transformer or foundation model architectures. Middle lane: an agentic layer in which large language models orchestrate tool calls to encoders, fusion models and knowledge bases to generate structured reports and explanations. Bottom lane: translational path from data curation and model development through external validation and deployment to decision support and digital twin--enabled in silico trials.}
\label{fig:translational_roadmap}
\end{figure*}

Three priorities define the next phase: continuous spatial modelling, imaging-informed molecular phenotyping, and clinically validated integration with digital twins. Fig.~\ref{fig:translational_roadmap} outlines the pathway from imaging-anchored models to decision support and in silico trials.

\subsection{Spatial modelling and cross-modal generative models}

An important next step is to treat the heart as a continuous 3D object with coupled imaging and molecular fields. Spatial statistical models and Gaussian processes defined on anatomical domains can capture smooth gradients of gene expression and imaging phenotypes, enabling interpolation between sampled locations and joint analysis of structure and function. As spatial atlases grow, such models may link mechanical forces, perfusion, and molecular responses at higher resolution.

Conditional cross-modal generative models could simulate plausible molecular profiles from imaging and vice versa, supporting uncertainty-aware imputation and counterfactual experiments. Incorporating physiological constraints and mechanistic priors remains an open challenge but would improve realism and trustworthiness.

\subsection{Imaging-informed molecular phenotyping and foundation models}

A long-term goal is to infer coarse molecular phenotypes from routine imaging with calibrated uncertainty. Achieving this at scale likely requires foundation models that combine large imaging archives with smaller paired imaging--omics cohorts. Self-supervised pretraining on millions of cardiac images followed by fine-tuning on imaging--transcriptomic pairs could yield representations aligned to molecular programmes. In parallel, single-cell foundation models trained on tens of millions of cells \cite{singlecell_foundation_2024,nicheformer_2025} provide transferable molecular encoders that can plug into cardiovascular fusion pipelines as atlas resources expand.

\subsection{Digital twins and integrative simulation}

Imaging-anchored multiomics is also converging with cardiac digital twins. Personalized computational models of anatomy and physiology derived from imaging can be augmented with patient-specific molecular context to simulate disease progression and therapy response \cite{corralacero2020_precision_cardiology,coorey2022_health_digital_twin,prakosa2018personalized}. Conversely, digital twin simulations can generate synthetic multimodal data for rare scenarios and support in silico trials \cite{sel2024_building_digital_twins,akbarialiabad2025_insilico_trials}. A complementary imaging-anchored example builds personalized 3D electrophysiological heart models directly from late-gadolinium-enhancement CMR: applied to post-infarction patients, these virtual-heart models simulate the propensity of each patient's reconstructed myocardium to sustain re-entrant arrhythmia and stratify sudden-death risk more accurately than established clinical metrics \cite{arevalo2016arrhythmia}. Integrating such imaging-derived digital twins with imaging-anchored multiomics pipelines could enable in silico evaluation of how molecular subtypes identified by imaging--omics integration respond to specific interventions, bridging the gap between observational multiomics discovery and interventional evidence. More broadly, integrating data-driven and mechanistic models will require careful calibration and validation, but offers a path to multiscale precision medicine.

\section{Conclusion}

Imaging has long been the visual backbone of cardiovascular medicine. In the era of multiomics, it can also serve as the computational backbone: a spatial and temporal reference frame onto which transcriptomic and spatial transcriptomic data are mapped. This review has framed \emph{imaging-anchored molecular phenotyping} as a central bioinformatics problem and outlined key methodological components required to address it: modality-specific representation learning, multimodal fusion via factor models, graphs, transformers, and contrastive learning, and integrative pipelines for aligning imaging with bulk, single-cell, and spatial transcriptomics.

Radiogenomic and spatial transcriptomic studies in cardiology are beginning to show how imaging phenotypes---fibrosis patterns, strain abnormalities, and plaque morphologies---relate to specific cell types, pathways, and niches. Multimodal fusion methods provide cross-modal latent spaces where disease subtypes, mechanistic hypotheses, and candidate biomarkers can be explored systematically. At the same time, practical issues surrounding data quality, harmonization, batch effects, sample size, fairness, and reproducibility remain central and should be treated as core design constraints rather than afterthoughts.

Looking ahead, imaging-anchored multiomics is likely to evolve along three main trajectories: towards richer spatial models that integrate imaging and molecular fields, towards foundation models and generative approaches that scale molecularly informed imaging analysis to broader populations, and towards incorporation into digital twins and agentic systems that embed these capabilities into clinical workflows. Achieving this vision will require close collaboration between cardiologists, data scientists, biologists, and engineers, but the potential payoff is substantial: a transition from predominantly descriptive imaging towards a more mechanistic, molecularly guided understanding of cardiovascular health and disease.

\par\vspace{10bp}
\noindent\fbox{\begin{minipage}{\dimexpr\columnwidth-2\fboxsep-2\fboxrule\relax}
\fontsize{7.5bp}{11bp}\selectfont
\textbf{Key Points}
\begin{itemize}
\item Imaging-anchored multiomics learns joint representations linking cardiac imaging phenotypes to bulk, sc/snRNA-seq, and spatial transcriptomics.
\item Cardiac imaging provides a spatial phenotype that supports voxel/region alignment with molecular states and cell-type programmes.
\item Fusion methods can accommodate missing modalities and batch effects but remain constrained by small paired cohorts.
\item Applications include radiogenomics, virtual transcriptomics, disease subtyping, mechanism discovery, and biomarker prioritization for cardiovascular disease.
\end{itemize}
\end{minipage}}
\par\vspace{8bp}

\section{Conflicts of interest}

None declared.

\section{Funding}

This work was supported by the National Science and Technology Council, Taiwan (grant number NSTC114-2221-E-038-015).

\section{Data availability}

No new data were generated or analysed in this review. All datasets and software mentioned are from previously published studies or publicly available resources cited in the references.


\end{document}